%% file: MAIN_1Column.tex
\documentclass[journal,12pt,onecolumn,draftclsnofoot,]{IEEEtran}

\usepackage[colorlinks]{hyperref}

\IEEEoverridecommandlockouts
\usepackage{cite}

\usepackage{soul}

\usepackage{multirow}
\usepackage{soul,color}
\usepackage{graphicx}
\usepackage{dsfont}
\usepackage{subcaption}

\usepackage[utf8]{inputenc}
\usepackage{mathtools}
\usepackage[mathscr]{euscript}
\usepackage{amsmath}
\usepackage{amssymb}
\usepackage{amsfonts}
\usepackage{verbatim}
\usepackage[T1]{fontenc}
\usepackage[utf8]{inputenc}
\usepackage{latexsym}
\usepackage{enumerate}
\usepackage{indentfirst}
\usepackage{calligra}
\usepackage{ulem}

\usepackage{array}
\usepackage{float}
\usepackage{bbm}

\usepackage{geometry}
\usepackage{eurosym}

\usepackage{hyperref}

\usepackage{caption}

\usepackage{tikz}
\usepackage{pgfplots}

\pgfplotsset{compat=1.8}
\usepgfplotslibrary{statistics}
\pgfmathdeclarefunction{fpumod}{2}{%
        \pgfmathfloatdivide{#1}{#2}%
        \pgfmathfloatint{\pgfmathresult}%
        \pgfmathfloatmultiply{\pgfmathresult}{#2}%
        \pgfmathfloatsubtract{#1}{\pgfmathresult}%
        \pgfmathfloatifapproxequalrel{\pgfmathresult}{#2}{\def\pgfmathresult{5}}{}%
    }

\usepackage{tabularx}
\usepackage{flushend}
\usepackage{textcomp}
\usepackage{xcolor}
\usepackage{import}

\usetikzlibrary{trees,decorations,shadows}
\tikzset{level 1/.style={sibling angle=45,level distance=4mm}}
\usetikzlibrary{arrows.meta}
\usepackage{forest}
\usetikzlibrary{external}

\let\oldtikzexternalgetnextfilename\tikzexternalgetnextfilename \renewcommand{\tikzexternalgetnextfilename}[1]{\oldtikzexternalgetnextfilename{#1}\expandafter\tikzsetnextfilename\expandafter{#1}}

\usepackage{amsmath}\usepackage{amssymb,amsthm}
\usepackage{amsfonts}
\usepackage{pgfplotstable}
\usepackage[outline]{contour}
\contourlength{1.2pt}

\pgfplotsset{compat=1.13} 

\usetikzlibrary{spy}
\usetikzlibrary{calc}
\usetikzlibrary{fadings}
\usetikzlibrary{patterns}
\usetikzlibrary{shadows}
\usetikzlibrary{mindmap}
\usetikzlibrary{backgrounds}
\usetikzlibrary{shapes.symbols}
\usetikzlibrary{shapes.multipart}
\usetikzlibrary{shapes.geometric}
\usetikzlibrary{automata,positioning}
\usetikzlibrary{decorations.fractals} 
\usetikzlibrary{decorations.markings}
\usetikzlibrary{decorations.pathreplacing}
\usetikzlibrary{decorations.pathmorphing}
    
\usepackage{bm}

\usepackage{nomencl}
\makenomenclature
\tikzset{edge from parent/.style={segment angle=10,draw}}

\tikzset{
  my rounded corners/.append style={rounded corners=2pt},
}

\def\BibTeX{{\rm B\kern-.05em{\sc i\kern-.025em b}\kern-.08em
    T\kern-.1667em\lower.7ex\hbox{E}\kern-.125emX}}

\renewcommand{\nomgroup}[1]{%
     \ifthenelse{\equal{#1}{O}}{\item[\textit{Operators}]}{%
        \ifthenelse{\equal{#1}{I}}{\item[\textit{Indices}]}{%
            \ifthenelse{\equal{#1}{A}}{\item[\textit{Acronyms}]}{%
            `\ifthenelse{\equal{#1}{V}}{\item[\textit{Variables and parameters}]}{}}}}}
\usepackage{scalerel}
\usetikzlibrary{svg.path}
\definecolor{orcidlogocol}{HTML}{A6CE39}
\tikzset{
    orcidlogo/.pic={
        \fill[orcidlogocol] svg{M256,128c0,70.7-57.3,128-128,128C57.3,256,0,198.7,0,128C0,57.3,57.3,0,128,0C198.7,0,256,57.3,256,128z};
        \fill[white] svg{M86.3,186.2H70.9V79.1h15.4v48.4V186.2z}
        svg{M108.9,79.1h41.6c39.6,0,57,28.3,57,53.6c0,27.5-21.5,53.6-56.8,53.6h-41.8V79.1z M124.3,172.4h24.5c34.9,0,42.9-26.5,42.9-39.7c0-21.5-13.7-39.7-43.7-39.7h-23.7V172.4z}
        svg{M88.7,56.8c0,5.5-4.5,10.1-10.1,10.1c-5.6,0-10.1-4.6-10.1-10.1c0-5.6,4.5-10.1,10.1-10.1C84.2,46.7,88.7,51.3,88.7,56.8z};
    }
}

\newcommand\orcidicon[1]{\href{https://orcid.org/#1}{\mbox{\scalerel*{ \begin{tikzpicture}[yscale=-1,transform shape]
                \pic{orcidlogo};
                \end{tikzpicture}
            }{|}}}}

\begin{document}


\title{{\Huge{Perspectives on distribution network flexible and curtailable resource activation and needs assessment}}}


\author{Md~Umar~Hashmi,~\IEEEmembership{Member,~IEEE}~\orcidicon{0000-0002-0193-6703},~Arpan~Koirala,~\IEEEmembership{Graduate~Student~Member~IEEE}~\orcidicon{0000-0003-4826-7137},~Hakan~Ergun,~\IEEEmembership{Senior~Member,~IEEE}~\orcidicon{0000-0001-5171-1986}, and~Dirk~Van~Hertem,~\IEEEmembership{Senior~Member,~IEEE}~\orcidicon{0000-0001-5461-8891}
\thanks{Corresponding author email: mdumar.hashmi@kuleuven.be}
\thanks{Md U. Hashmi, A. Koirala, H. Ergun,  and D. Van Hertem are with KU Leuven, division Electa \& EnergyVille, Genk, Belgium}
\thanks{This work is supported by the H2020 EUniversal project, grant ID: 864334 (\url{https://euniversal.eu/}) and the energy transition funds project BREGILAB organized by the FPS economy, S.M.E.s, Self-employed and Energy.}}



\maketitle

\begin{abstract}

{A curtailable and flexible resource activation framework for solving distribution network (DN) voltage and thermal congestions is used to quantify three important aspects with respect to modelling low voltage networks.}
This framework utilizes the network states in the absence of such flexible or curtailable resources as the input for calculating flexibility activation signal (FAS). 
The FAS has some similarities with optimal power flow duals {associated with power balance constraint}.
FAS due to drooping design, {incentivize corrective flexibility activation}
prior to any network limit violations.
{The nonlinear resource dispatch optimal power flow (RDOPF) utilizes FAS for the activation of flexible and curtailable resources. Solving the OPF problem for a large system is computationally intensive, and second-order cone (SOC) relaxation is often applied in the literature.}
{First,} we highlight the multi-objective nature of SOC relaxed RDOPF. A Pareto front tuning mechanism {is proposed for choosing loss penalty factor} while reducing the optimality gap of the SOC relaxed RDOPF.
{Secondly, we} present a methodology for evaluating temporal and locational flexibility needs assessment of a DN, which DSO's can utilize for flexibility planning {in operational timescales and procurement in the flexibility market}.
{Lastly, we} quantify the impact of reactive power flexibility for a DN with varying load power factors.
{Numerical simulations indicate that the presence of reactive flexibility reduces the active power flexibility needs by 50\% for the test feeder with 0.8 aggregated load power factor.}
\end{abstract}

\begin{IEEEkeywords}
Flexibility assessment, Load curtailment, Optimal power flow, Convex optimization, Pareto optimal, Reactive power
\end{IEEEkeywords}

\pagebreak

\tableofcontents

\pagebreak

\begin{table}[!htbp]
\small
\begin{tabular}{ll}
\normalsize{\textcolor{black}{Abbreviation}} &                                \\
DN           & Distribution network           \\
DSO          & Distribution system operator   \\
FAS          & Flexibility activation signal  \\
HVAC & Heating, ventilation, and air conditioning\\
LMP          & Locational marginal price      \\
NVS          & Nodal voltage sensitivity      \\
OPF          & Optimal power flow             \\
RDOPF        & Resource dispatch OPF          \\
SOC          & Second order cone             
\end{tabular}
\end{table}

\section{Introduction}
Traditionally, the distribution system operator (DSO) relied on a fit-and-forget \textcolor{black}{approach for designing the network.}
However, there will be a greater need to integrate flexible and curtailable resources 
to cope with growing distributed generation (DG) installations and new consumption patterns \textcolor{black}{while ensuring network integrity}. 
Distribution network (DN) vulnerabilities, \textcolor{black}{such as} voltage and thermal limit violations, often require resource activation in the vicinity. 
{Flexible and curtailable resource planning will be crucial for reliable DN operation in presence of large amounts of demand and distributed generation fluctuations, which could probably cause DN voltage and/or thermal \textcolor{black}{overload} issues.
Several DSOs, such as Mitnetz Strom in Germany \cite{mitnetz}, and electricity retailers, such as Luminus in Flanders, Belgium \cite{luminus}, already promote interruptible installations \textcolor{black}{of flexible loads such as water heaters}.
Accenture's digitally enabled grid shows the future need for flexibility in DN \cite{accenture_blog}.
The H2020 InterFlex project demonstrates the flexibility activation and market design using four demonstrations in Germany (Avacon), Sweden (E.ON), France (Enedis, ENGIE, and EDF), and the Netherlands (Enexis, Jedlix, TNO, and Sympower) \cite{dumbs2019market, interflexpr}.
{On similar lines, the H2020 EUniversal project aims to create a universal and reusable approach for European partners for using flexible resources via dedicated market design for mitigating network issues through three demonstrations in Germany, Poland and Portugal \cite{euniversal}.}
Report \cite{radecke2019markets} discusses the European landscape for the execution of local markets for flexible resources. They observe market design for incentivizing flexible resources for providing grid services as a major challenge.}
\textcolor{black}{The increased relevance of DN flexibility motivates our present work}.



Authors in \cite{meibetaner2019co} present a case study of Northern Germany, where every MW of additional wind installation will require 0.7 MW of flexibility for avoiding curtailment of renewable generation.
Prior works \cite{fonteijn2018flexibility,tsaousoglou2021mechanism,torbaghan2016local,laur2020optimal} present market mechanisms for end-user flexible resources for solving low voltage network issues.
The market mechanism design presented in \cite{fonteijn2018flexibility} utilizes contractual flexibility by DSOs for H2020 Interflex project.
Authors in \cite{tsaousoglou2021mechanism} propose an efficient market design for flexible resources. They observe that strategic misreporting of resources could lead to market inefficiency.
Authors in \cite{torbaghan2016local} propose a hierarchical, bi-level optimization problem for facilitating optimal bidding of flexible resources for DSO in day-ahead and real-time timeframes.
Authors in \cite{agbonaye2021mapping} observe that locational temporal flexibility mapping is crucial for avoiding future network issues. 

DN fluctuations due to volatility in distributed demand and/or generation need to be \textcolor{black}{amortized to permissible levels}.
Authors in \cite{ma2013evaluating}, developed a business case for the growth of such flexible resources by appropriately designing electricity markets that promote the growth of flexible resources while ensuring their profitability.
In this work, we propose a hierarchical, flexible and curtailable resource activation mechanism \textcolor{black}{for a DSO through which they can quantify the flexibility needs of a DN}. Resources participating in such a market have a lower temporal priority such as water heaters, HVAC systems, thermostatic loads, heat pumps, batteries etc \cite{chen2018distributed}. 
Authors in \cite{ramos2016realizing} observe that a location specific flexibility market is crucial for procuring responsive resources for various grid services.
We propose a location-aware resource activation mechanism for DNs.
%
We utilize nodal voltage sensitivity (NVS) in order to bring the locational aspect in the flexibility activation signal (FAS) design.
NVS measures the impact of per unit change in active and reactive power on the voltage change at that node and the rest of the DN. 
Voltage sensitivities are widely used in many power system applications, such as
battery management \cite{hashemi2014scenario, giannitrapani2016optimal, shafiq2019optimal},
inverter operation \cite{zad2018new, demirok2011local, zhang2017novel, tao2018voltage},
on-line load sensitivity \cite{de2016line}.
The voltage sensitivity matrix can be calculated using (a) Jacobian-matrix inverse based on linearized power flow equations used in the Newton-Raphson power flow, (b) perturb and observe method, (c) admittance compound matrix \cite{christakou2013efficient}, (d) fitting-function based sensitivity approach \cite{zhang2017novel}.
{This work uses \textit{perturb-and-observe} to approximate NVS. }



\begin{figure*}[!htbp]
	\center
	\includegraphics[width=0.68\linewidth]{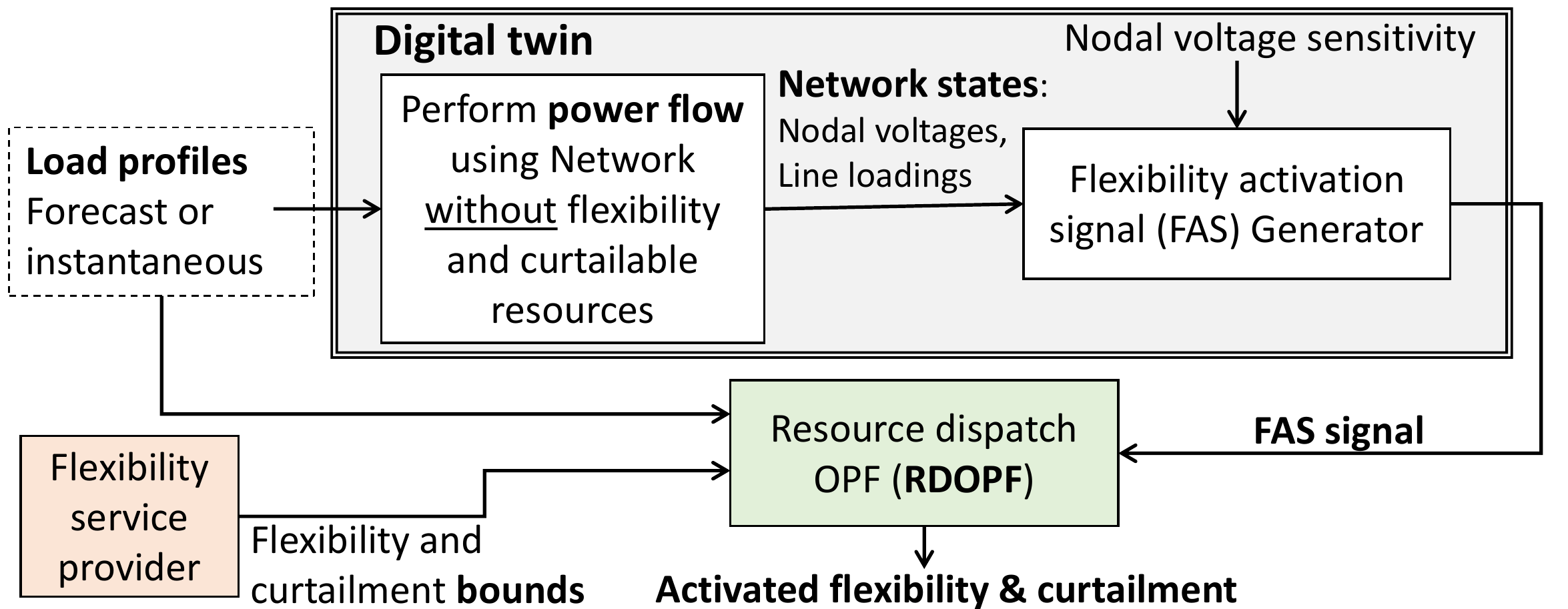}
	\vspace{-2pt}
	\caption{\small{\textcolor{black}{RDOPF with a digital twin based FAS generation.}}}
	\label{fig:pic88}
\end{figure*}

{FAS design takes as input NVS and the forecasted or measured state of the network, which includes nodal voltage magnitude, and thermal loading of branches, see Fig. \ref{fig:pic88}.}
The FAS design is motivated by volt-watt and volt-Var inverter control. Volt-watt and volt-Var
inverter control in standalone or in combination are popular \cite{weckx2014combined,karagiannopoulos2017hybrid, https://doi.org/10.48550/arxiv.2207.10248}. 
\textcolor{black}{These inverter control policies use a permissible limit, beyond which it provides
droop based correction of local voltage.}
Similar to these inverter control policies, FAS is active in case DN parameters exceed a permissible level, thus incentivizing resource activation.

We observe that the proposed FAS design has some similarities with optimal power flow (OPF) duals, often used as locational marginal prices (LMP) \cite{conejo2005locational}. 
\textcolor{black}{The duals of OPF constraints, collectively utilized as distribution LMPs (DLMPs) for providing different services such as congestion management, power balancing for active and reactive power and loss minimization \cite{bai2017distribution}. 
DLMPs are used for congestion management in prior works such as \cite{li2013distribution, zhao2019congestion}.
In this work, we will refer to the dual variable associated with the power balance constraint as LMP.}
The proposed FAS provides more information than the OPF duals due to the droop based design. 
\textcolor{black}{
On the contrary, the OPF duals are only active (non-zero) in case of network constraint violations {due to the objective function structure, which considers flexibility activation cost while the generation ramping constraint, the generation cost and DN loss penalty are not considered in the objective function}.} 
Although, FASs are analogous to LMPs, we consider only the variable component of such LMPs. In reality, DSOs may also need to settle the capacity cost even if the flexibility is not utilized. The capacity cost itself is not considered in this work.
{Furthermore}, the market operator \textcolor{black}{or an aggregator} may need to create a framework to measure the performance of flexible resources when requested to activate, similar to an ancillary services market operated under PJM \textcolor{black}{Interconnection} in the US \cite{pjm}.
In this work, we assume full performance by resources when activated, depending on their reported limits.



The proposed resource dispatch optimal power flow (RDOPF) problem is convexified using second order cone (SOC) relaxations \cite{hashmi2021sest, jabr2006radial,molzahn2019survey}.
Authors in \cite{gan2012exactness} observe that convex relaxations for OPF problems are exact provided there are no DGs which could lead to reverse power flow and the R/X ratio is non-decreasing from substation to the end of DN. The first condition may not be valid for DNs with a large amount of DG generation sources and flexible load/generation.
Thus, we \textcolor{black}{assume} that convex relaxations for RDOPF are not exact. 

{The reactive compensation is crucial for the healthy operation of the transmission networks and is traditionally ignored for DNs. However, with growing inverter interfaced DGs and new loads, \textcolor{black}{the DN power factor is observed to be deteriorating for the island of Madeira} \cite{hashmi2020arbitrage}.
This degrading DN power factor leads to reduced supply efficiency and, if left unchecked, could hurt the \textcolor{black}{financial sustainability} of DSOs and power suppliers.
Taking this aspect into consideration, utilities in France \cite{linkyfrance}, Germany \cite{europepf}, and Uruguay \cite{uruguay} are some countries where prosumers are penalized for poor power factor, thus incentivizing reactive power compensation.
From {the} system operator's perspective, regulated reactive power leads to {a} reduction in DN losses \cite{alkaabi2018short}.}
\textcolor{black}{Reactive power flexibility is growing in importance in active distribution network operation {\cite{brandao2019optimal,camacho2014active}}.}

{The starting point of this work is to use the information provided by an intermediary between flexibility owners and the DSO, referred to as the flexibility service provider (FSP).
The FSP provides the flexibility's operational constraint information, such as maximum and minimum power limits and energy limits.}
Similar to \cite{torbaghan2019optimal}, this paper assumes that instantaneous flexibility, in the form of lower and upper envelopes, is known to DSO, e.g., provided by an intermediary actor as a resource aggregator. Here we focus on flexible and subsequently curtailable resource activation using the proposed flexibility activation priority design, {which considers} the nodal voltage sensitivity, local voltage measurement and connected line {currents}. 
\textcolor{black}{The mechanism of FAS calculation using a digital twin which does not consider flexible and curtailable resources is fed to RDOPF, see Fig. \ref{fig:pic88}}.

\textcolor{black}{
The proposed RDOPF framework can be used by DSOs for\\
$\bullet$ \textit{Real-time application}: activation of controllable resources in DN providing flexibility and curtailment,\\
$\bullet$ \textit{Time-ahead application}: quantifying DN flexible and curtailable resource needs; such as day-ahead planning,\\
$\bullet$ \textit{Valuation of resources}:} 
The flexibility activation priority might be
cleared in the energy market in future, and the DSO can take advantage of this framework for valuing their flexible resources in market-clearing.

Note that with local measurement of line currents and nodal voltages used in the proposed activation signal design, it can be used as a {distributed control without communicating the FAS}, which will be the future direction of this work.
\vspace{-1pt}

\subsection{Contributions of the paper}

\textcolor{black}{This work is an extension of prior work presented in \cite{hashmi2021sest}, where we propose the RDOPF framework. 
In this work, this framework is extended for addressing three main points of attention with respect to \cite{hashmi2021sest}.\\
$\bullet$ Firstly, we observe that SOC formulation as used in \cite{hashmi2021sest} leads to excessive load curtailment compared to generation curtailment in contrast to the nonlinear, nonconvex formulation of the RDOPF. In this work, the gap between SOC and nonlinear formulation of RDOPF is reduced by the introduction of a loss penalty in the objective function. We propose a Pareto optimal mechanism to tune the loss penalty factor. This will avoid the
\textcolor{black}{excess flexibility activation}
for the sake of loss minimization.\\
$\bullet$ Secondly, we provide a framework to perform a flexibility needs assessment for a DN. Metrics are proposed for quantifying temporal and locational flexibility and curtailment needs of the DN.
{It is observed that a DN's flexibility needs depend on the valuation mechanism of different resources, and a flat activation cost leads to minimum activation of flexibility.}
\textcolor{black}{DSO's can utilize flexibility needs for flexibility procurement in operational timescales and flexibility planning in longer timescales.}\\
$\bullet$ Thirdly, we quantify the impact of reactive power flexibility on the distribution network’s active power flexibility needs, as well as the total flexibility activation cost in dependence of the load power factor. We show that the active power flexibility need and DSO profits are negatively affected by decreasing load power factors, demonstrating the value of reactive power flexibility. 
}



The paper is organized as follows. 
Section~\ref{section2} presents the nodal voltage sensitivity calculations for active and reactive power perturbations and the model used for calculating the flexibility activation signal.
Section~\ref{section4} formulates the optimization problem for resource activation and also provides SOC relaxations for the RDOPF.
Section \ref{section5} presents the three numerical case studies, and
section~\ref{section6} concludes the paper.

\pagebreak




\section{Flexibility activation signal design}
\label{section2}
The activation of flexible resources is based on the instantaneous network state. 
Motivated by \textcolor{black}{the capabilities provided by} volt-var and volt-watt inverter control, the design for FAS is proposed in \cite{hashmi2021sest} and summarized here in this work.
The locational aspect of such resources are brought in by utilizing the NVS. NVS indicate the sensitivity of a network node towards active and reactive power perturbation on nodal voltage magnitude. 
The mechanism used to calculate NVSs is detailed in Section \ref{sec:nvs}.
The FAS design generates a nodal activation signal for flexible resources. DN parameters, such as line currents, and line loadings, are not nodal. A branch to nodal projection mechanism is detailed in 
our prior work \cite{hashmi2021sest}.
The FAS design detailed in Section \ref{sec:fasdesign} considers the instantaneous nodal voltages and projected nodal line loadings. 
\vspace{-3pt}
\input{fig_ref}
\vspace{-10pt}

\subsection{Notation}
A DN consists of nodes,
branches, loads and generators. 
A DN is characterized by $<\mathscr{N},E>$, where $\mathscr{N}$ denotes nodes and $E$ denotes branches connecting a pair of nodes.
Each node $i \in \mathscr{N}$ and time $t$ has two variables, i.e., voltage magnitude ($V_{i,t}$) and phase angle ($\theta_{i,t}$) which are governed by power injection and load magnitude. The branch admittance {$Y_{(i,j)}$},  governs the flow and losses \textcolor{black}{between nodes $i$ and $j$, where $(i,j) \in E$}.
Nodes with loads connected are denoted as $\mathscr{N}_L \subset \mathscr{N}$. These nodes have active and reactive power loads denoted as $P^d_{i,t}$ and $Q^d_{i,t}$.
Nodes with generators connected are denoted as $\mathscr{N}_G \subset \mathscr{N}$, have active and reactive power generation denoted as $P^g_{i,t}$ and $Q^g_{i,t}$. 
{The operator $\mathbbm{1}(.)$ returns 1 if the condition is true and 0 otherwise.} 
\subsection{Nodal voltage sensitivity}
\label{sec:nvs}
The voltage sensitivity of a node $i$ towards active and reactive power load perturbations at node $j$ for scenario $k$ is given as
\begin{equation}
    \Psi_{j,i}^k = \Big| \frac{V_i^k - V_i^{k_0}}{P_j^k - P_j^{k_0}}\Big|,~~\beta_{j,i}^k = \Big|\frac{V_i^k - V_i^{k_0}}{Q_j^k - Q_j^{k_0}}\Big|,~ i,j \in \mathscr{N}.
\end{equation}
where the active load at a node $j$ is modified from $P_j^{k_0}$ ($Q_j^{k_0}$) to $P_j^{k}$ ($Q_j^{k}$), due to which the new power flow results in a change in voltage  at each node $i \in \mathscr{N}$. $V_i^{k_0}$ denotes the voltage at node $i$ corresponding to load $P_j^{k_0}$ ($Q_j^{k_0}$).
The NVS for a node $x$ are calculated based on $U$ number of power flow calculations performed at different active and reactive power loadings and are given as
\begin{gather}
    \Psi_{x} = \frac{1}{U} \sum_{k=1}^{U} \sum_{i=1}^N \Psi_{x,i}^k,~~
    \beta_{x} = \frac{1}{U} \sum_{k=1}^{U} \sum_{i=1}^N \beta_{x,i}^k.
    \label{eq:sensitivityMatP}
\end{gather}
{The study in}~\cite{vanet2015sensitivity} indicates that higher network loading implies greater voltage sensitivity. In order to eliminate the loading effects, $\Psi_{x} $ and $\beta_{x}$ are calculated at different loading conditions and averaged over \textcolor{black}{one hundred} \textcolor{black}{Monte Carlo} simulations. 



\subsection{Flexibility activation signal design}
\label{sec:fasdesign}
Activating flexible resources is crucial for the healthy operation of DNs with DGs and new loads, which will be more prone to thermal violations, under-voltage and over-voltage phenomenons. 
The flexibility activation considers the locational aspect of flexible resources for fairly valuing for their responsive services.
Further, flexible resources need to be activated only when the network state is {near or above the} voltage or thermal {limits} 
at one or more nodes. 
Considering these aspects, we propose a FAS design combining NVS shown in \eqref{eq:sensitivityMatP}  with instantaneous network states. 
These sensitivity values are analogous to merit-order for flexibility activation. In a radial DN, the sensitivity at the end of the feeder will be significantly higher than the voltage at the beginning of the feeder. 
\textcolor{black}{Fig. \ref{fig:pic88} shows the digital twin used for generating network states in absence of flexible and curtailable resources are used as an input for FAS generation.}
The FAS design {has} two nodal signals for active and reactive power flexible resources. Each signal consists of voltage and thermal components, see Fig.~\ref{fig:fasdesign}. Voltage component is decided based on nodal voltage magnitude. If the nodal phase voltage exceeds $1+\Delta V_{\text{perm}}$ or dips below $1-\Delta V_{\text{perm}}$, the FAS voltage component is non-zero and promotes the flexibility activation at that node.
Similarly, the thermal component of FAS is non-zero if projected thermal loading\footnote{\textcolor{black}{Thermal loading refers to the percentage of branch capacity being used. Lines are congested if thermal loading exceeds 100\% of the line capacity.}} exceeds $\Delta T_{\text{perm}}$ or dips below $-\Delta T_{\text{perm}}$ thresholds.
\textcolor{black}{The projected thermal loading is introduced in \cite{hashmi2021sest}, which proposes a mechanism to convert branch thermal loading into nodal signal.}
The FAS follows a drooping behavior similar to volt-watt and volt-var inverter control \cite{weckx2014combined}, \cite{karagiannopoulos2017hybrid}.
The upper bound of the voltage component of the active power FAS is denoted as $\text{VC}^{\max}_{i,P}$ and the thermal component is denoted as $\text{TC}^{\max}_{i,j,P}$. 
These saturating levels are governed by nodal voltage sensitivities calculated in \eqref{eq:sensitivityMatP} and given as {
\begin{equation}
     \text{VC}^{\max}_{i,P} = f_P(\Psi_i),~
    \text{TC}^{\max}_{i,j,P} = g_P(\beta_i),~ i \in \mathscr{N}.
    \label{eq:sensitivityComponetFlexPriceP}
\end{equation}
}
Similarly, the reactive component of the flexibility value due to voltage and thermal limit violation are denoted as {
\begin{equation}
    \text{VC}^{\max}_{i,Q} = f_Q(\Psi_i),~
    \text{TC}^{\max}_{i,j,Q} = g_Q(\beta_i),~ i \in \mathscr{N}.
    \label{eq:sensitivityComponetFlexPriceQ}
\end{equation}}
The FAS for active power flexibilities are denoted as
\begin{subequations}
\label{eq:flexibilityPrices}
\label{eq:costflex}
\small{
\begin{equation} \begin{split}
    &\lambda_{i,t}^{\text{flex}_P-} =   
    \mathbbm{1}\Big(V_{i,t}\geq V_{\max}\Big)(-\text{VC}^{\max}_{i,P}) +
    \mathbbm{1}\Big(T_{i,j,t} \geq 100\Big)(-\text{TC}^{\max}_{i,j,P})  \\&+
    \mathbbm{1}\Big(V_{i,t}\in (1+ \Delta V_{\text{perm}}, V_{\max})\Big) \frac{(-\text{VC}^{\max}_{i,P}) \Big(V_{i,t} - (1+\Delta V_{\text{perm}})\Big)}{\Big(V_{\max} - ( 1+ \Delta V_{\text{perm}})\Big)} \\& +
    \mathbbm{1}\Big(T_{i,j,t} \in (\Delta T_{\text{perm}}, 100) \Big)
    \frac{(-\text{TC}^{\max}_{i,P}) \Big(T_{i,t} - \Delta T_{\text{perm}}\Big)}{\Big(100- \Delta T_{\text{perm}}\Big)},
\end{split} \end{equation} 
\vspace{-10pt}
\begin{equation} \begin{split}
    &  \lambda_{i,t}^{\text{flex}_P+} =
 \mathbbm{1}\Big(V_{i,t}\leq V_{\min}\Big)\text{VC}^{\max}_{i,P} + 
    \mathbbm{1}\Big(T_{i,j,t} \geq 100\Big)\text{TC}^{\max}_{i,j,P} + \\ & 
    \mathbbm{1}\Big(V_{i,t}\in (V_{\min}, 1- \Delta V_{\text{perm}})\Big) \frac{\text{VC}^{\max}_{i,P} \Big(V_{i,t} - (1-\Delta V_{\text{perm}})\Big)}{\Big(V_{\min} - ( 1- \Delta V_{\text{perm}})\Big)} +\\&
    \mathbbm{1}\Big(T_{i,j,t} \in (\Delta T_{\text{perm}}, 100) \Big)
    \frac{\text{TC}^{\max}_{i,P} \Big(T_{i,t} - \Delta T_{\text{perm}}\Big)}{\Big(100- \Delta T_{\text{perm}}\Big)}.
\end{split} \end{equation}
} \end{subequations}
\vspace{-1pt}
The FAS for reactive power, $\lambda_{i,t}^{\text{flex}_Q+}$ and $\lambda_{i,t}^{\text{flex}_Q-}$, can be derived similarly as shown in \eqref{eq:costflex}. Note that $\lambda_{i,t}^{\text{flex}_P+}$ is associated with $\Delta P_{i,t}^{\text{flex}+}$ and so on.

\pagebreak

\section{Optimization for resource activation}
\label{section4}
In this section, we detail the resource dispatch optimal power flow (RDOPF) proposed in \cite{hashmi2021sest}.
The DSO activates distributed flexible and curtailable resources to avoid DN voltage, and thermal issues, which otherwise could have happened. 
The DSO aims to minimize the activation cost.
The bounds for flexible and curtailable resources are assumed to be known. Next we detail the constraints and the optimization formulation.

\subsection{Flexibility definition}
The flexible resources are defined based on the ramp up and down, active (P) and reactive (Q) power levels. The ramp-up P flexibility increases nodal load and vice versa.
Similarly, Q flex is defined as capacitive and inductive.
These variables are 
\begin{IEEEeqnarray}{llll}
\IEEEyesnumber \label{eq:flexibility} 
    \Delta P^{\text{flex}+}_{ i,t} & \in & [0, P^{\text{flex}}_{\max, i,t}], & (\text{P injection or ramp down}),\\
    \Delta P^{\text{flex}-}_{ i,t} & \in & [P^{\text{flex}}_{\min, i,t}, 0],  & (\text{P consumption or ramp up}),\\
    \Delta Q^{\text{flex}+}_{ i,t} & \in & [0, Q^{\text{flex}}_{\max, i,t}],  & (\text{Q injection}),\\
     \Delta Q^{\text{flex}-}_{ i,t} & \in & [Q^{\text{flex}}_{\min, i,t}, 0],  & (\text{Q consumption}).
\end{IEEEeqnarray}
Thus, $\Delta P^{\text{flex}}_{ i,t} = \Delta P^{\text{flex}+}_{ i,t} + \Delta P^{\text{flex}-}_{ i,t}$ and $\Delta Q^{\text{flex}}_{ i,t} = \Delta Q^{\text{flex}+}_{ i,t} + \Delta Q^{\text{flex}-}_{ i,t}$.
Each flexibility component for ramp up and ramp down P and Q have an associated cost value denoted as $\lambda_{ i,t}^{\text{flex}_P+} \geq 0,\lambda_{ i,t}^{\text{flex}_P-} \leq 0, \lambda_{ i,t}^{\text{flex}_Q+} \geq 0,\lambda_{ i,t}^{\text{flex}_Q-} \leq 0$ respectively.
For cases where $\lambda_{ i,t}^{\text{flex}_P+},\lambda_{ i,t}^{\text{flex}_P-}$ are both zero implying voltage and line loadings are within permissible bounds, therefore, the min and max ranges for $\Delta P_{ i,t}^{\text{flex}+}$ and $\Delta P_{ i,t}^{\text{flex}-}$ should be zero. 
Similarly, for cases where $\lambda_{ i,t}^{\text{flex}_Q+},\lambda_{ i,t}^{\text{flex}_Q-}$ are both zero, the min and max ranges for $\Delta Q_{ i,t}^{\text{flex}+}$ and $\Delta Q_{ i,t}^{\text{flex}-}$ should be zero. 
In the absence of the above conditions, the power balance constraint in RDOPF implementation will not accurately represent the DN, as numerical optimization could trigger flexibilities even when no network issues are present. This problem can be solved by introducing an integer variable in the RDOPF or by redefining the flexibility constraint in \eqref{eq:flexibility} as
\begin{IEEEeqnarray}{lllll}
\IEEEyesnumber\label{eq:flexibility2} 
    \Delta P^{\text{flex}+}_{ i,t} &\in& [0, z_{i,t}^{s1} P^{\text{flex}}_{\max, i,t}] &=& [0,  P^{\text{flex}N}_{\max, i,t}],\\
    \Delta P^{\text{flex}-}_{ i,t} &\in& [z_{i,t}^{s2} P^{\text{flex}}_{\min, i,t}, 0] &=& [ P^{\text{flex}N}_{\min, i,t}, 0], \\
    \Delta Q^{\text{flex}+}_{ i,t} &\in& [0, z_{i,t}^{s3} Q^{\text{flex}}_{\max, i,t}] &=& [0,  Q^{\text{flex}N}_{\max, i,t}], \\
     \Delta Q^{\text{flex}-}_{ i,t} &\in& [z_{i,t}^{s4} Q^{\text{flex}}_{\min, i,t}, 0] &=& [ Q^{\text{flex}N}_{\min, i,t}, 0],
\end{IEEEeqnarray}
where $z_{i,t}^{s1}, z_{i,t}^{s2}, z_{i,t}^{s3}, z_{i,t}^{s4}$ denotes binary variables. These binary variables are calculated as
$
    z_{i,t}^{s1} = \mathbbm{1}( \lambda^{\text{flex}P+}_{ i,t} \ne  0),
    z_{i,t}^{s2} = \mathbbm{1}( \lambda^{\text{flex}P-}_{ i,t} \ne  0), 
    z_{i,t}^{s3} = \mathbbm{1}( \lambda^{\text{flex}Q+}_{ i,t} \ne  0), 
    z_{i,t}^{s4} = \mathbbm{1}( \lambda^{\text{flex}Q-}_{ i,t} \ne  0).
$
Since the FASs are calculated prior to solving the resource dispatch optimization problem, 
thus flexibility limits calculated as in \eqref{eq:flexibility2} avoid the use of binary variables in the RDOPF.

\subsection{Load and generation curtailment}
The load and generation curtailment cost is set at a level higher than the highest value of the FAS. This will ensure curtailment of load and generation are avoided if flexibility activation can solve network congestion.
In order to ensure a feasible solution of RDOPF, the limits for generation and load curtailment are defined as
\begin{equation}
\Delta P^{G}_{ i,t} \in[0, P^{g}_{ i,t}], ~~~~
\Delta P^{\text{curt}}_{ i,t}  \in [0, P^{d}_{i,t}],
  ~\forall i \in \mathscr{N_L}.
  \label{eq:gencurt}
\end{equation}

\subsection{Optimization formulation}
The decision variables for the optimization are $\Gamma = \{P_{j,t}^g,\Delta P^{\text{flex}}_{i,t}, \Delta Q^{\text{flex}}_{i,t},$ $ \Delta P^{\text{curt}}_{i,t}, \Delta P^{\text{G}}_{i,t} \}$ which denote active power flexible resource activated, reactive power flexible resource activated, active power generation curtailment and load shedding, respectively.
The objective function is given as
\begin{equation}\begin{split}
\label{eq:objectivefun}
   & \sigma(\rho, \lambda_{i,t}^{\text{flex}}, \lambda_{i,t}^{\text{curt}_P}, \lambda_{i,t}^{\text{curt}_G} ) =  
    \lambda_{\text{loss}} \rho(t) + \\& 
     \lambda_{i,t}^{\text{flex}_P+} \sum_{i \in \mathscr{N} }\Delta P^{\text{flex}+}_{i,t} + 
     \lambda_{i,t}^{\text{flex}_P-} \sum_{i \in \mathscr{N} }\Delta P^{\text{flex}-}_{i,t}    +
     \lambda_{i,t}^{\text{flex}_Q+} \sum_{i \in \mathscr{N} }\Delta Q^{\text{flex}+}_{i,t} \\ & +
     \lambda_{i,t}^{\text{flex}_Q-} \sum_{i \in \mathscr{N} }\Delta Q^{\text{flex}-}_{i,t} + 
     \lambda_{i,t}^{\text{curt}_G} \sum_{i \in \mathscr{N} }\Delta P^{\text{G}}_{i,t}+ 
    \lambda_{i,t}^{\text{curt}_P} \sum_{i \in \mathscr{N} }\Delta P^{\text{curt}}_{i,t}, \vspace{-6pt}
\end{split}\end{equation}
where 
$\rho$ denotes line losses and $\lambda_{\text{loss}}$ denotes the penalty associated with line losses.
We can select the objective function parameter values as 
$
   0 \leq \max(\lambda_{i,t}^{\text{flex}_P+}, |\lambda_{i,t}^{\text{flex}_P-}|)
   <  \lambda_{i,t}^{\text{curt}_G}, \lambda_{i,t}^{\text{curt}_P}.
    \label{eq:condition}
$
The cost of generation and load curtailment is set higher than the highest flexibility activation priority levels. 
The above condition ensures that no load shedding or generation curtailment are performed before activation of available flexible resources. 
The full nonlinear optimization formulation (RDOPF) is 
\begin{subequations}
\begin{equation}\begin{split}
\label{eq:porgobjective}
  (P_{\text{org}})~~ \underset{\substack{\Gamma}}{\text{min}}~ \sum_t \sigma(P_{j,t}^g, \rho, \lambda_{i,t}^{\text{flex}}, \lambda_{i,t}^{\text{curt}_P}, \lambda_{i,t}^{\text{curt}_G} ) 
\end{split}\end{equation} 
\text{subject to,} \eqref{eq:flexibility2}, \eqref{eq:gencurt} and
\begin{equation}
   ~V_{\min} \leq |V_{i,t}| \leq V_{\max}, ~ \forall i \in \mathscr{N} , t \in \{1,..,T\},
  \label{eq:voltage}
\end{equation}
\begin{equation}\begin{split}
 &  (P^g_{i,t} - \Delta P^G_{i,t}) - (P^d_{i,t} - \Delta P^{\text{curt}}_{i,t} - \Delta P^{\text{flex}}_{i,t}) +\\ & j(Q^d_{i,t}  - \Delta Q^{\text{flex}}_{i,t}) = \sum s_{i,j,t},  ~~\forall~ i,j \in \mathscr{N} ,
 \label{eq:powerbalance}
\end{split}\end{equation}
\begin{equation}
  |S_{i,j, t}| < s_{i,j}^{\max}, ~~\forall~ i,j \in \mathscr{N}     \label{const:thermal}, 
\end{equation}
\begin{equation}
  P^g_{i,t} \in [P^g_{\min,i},P^g_{\max,i}] , ~~\forall~ i \in \mathscr{N_G}     \label{const:genlimit}, 
\end{equation}
\begin{equation}
  S_{i,j,t}  = \textbf{Y}_{i,j}^*V_{i,t}V_{i,t}^* - \textbf{Y}_{i,j}^*V_{i,t}V_{j,t}^*, ~ \forall (i,j) \in E \cup E^R, 
  \label{eq:ohmslaw}
\end{equation}
\begin{equation}
  \angle (V_{i,t} V_{j,t}^*) \in [\theta_{i,j}^{\min}, \theta_{i,j}^{\max}], ~~\forall~ i,j \in \mathscr{N}, 
  \label{const:phase}
\end{equation}
\end{subequations}
\eqref{eq:voltage}, \eqref{const:thermal} and \eqref{const:phase} denote the voltage constraint for nodes, thermal constraint and phase angle constraints for branches, respectively. \eqref{eq:powerbalance} denotes the nodal balance of active and reactive power in the network.
\eqref{const:genlimit} denotes the generator output power limits.
\eqref{eq:ohmslaw} denotes Ohm's law. Flexibility limits for active and reactive ramp up and ramp down are denoted in
\eqref{eq:flexibility2}.
\eqref{eq:gencurt} place limits on generation and load curtailment, respectively.

\subsection{Convexification of optimization}
Optimization formulation $P_{\text{org}}$ is non-convex due
to
\eqref{eq:voltage}.
The voltage constraint in the power flow equations causes $P_{\text{org}}$ to be nonlinear. 
Second order cone (SOC) relaxation according to \cite{jabr2006radial} is used to convexify the problem.
This formulation is referred to as SOC RDOPF:
\begin{subequations}
\label{eq:socrelax}
\begin{equation}
    |W_{i,j,t}|^2  \leq W_{i,i,t} W_{j,j,t}, \text{ where } V_{\min}^2 \leq W_{i,i,t}\leq V_{\max}^2,
\end{equation}
\begin{equation}
    S_{i,j,t}  = \textbf{Y}_{i,j}^*W_{i,i,t} - \textbf{Y}_{i,j}^*W_{i,j,t}, ~ \forall (i,j) \in E \cup E^R
\end{equation}
\end{subequations}

The convex formulation of $P_{\text{org}}$ is denoted as
\begin{gather*}
    (P_{\text{cvx}})~~ \text{objective function~~~~~~~Eq.}~\ref{eq:porgobjective},\\
    \text{subject to ,}~
    \eqref{eq:flexibility2}, \eqref{eq:socrelax}, \eqref{eq:powerbalance}, \eqref{const:thermal}, \eqref{const:phase}, \eqref{const:genlimit},
    \eqref{eq:gencurt}. 
\end{gather*}

\pagebreak

\section{Numerical studies}
\label{section5}
The network details for which numerical results are presented are detailed below.
The numerical simulations are performed using PowerModels.jl in Julia / JuMP
\cite{coffrin2018powermodels}.

\begin{figure}[!htbp]
    \centering
    \includegraphics[width=0.7\linewidth]{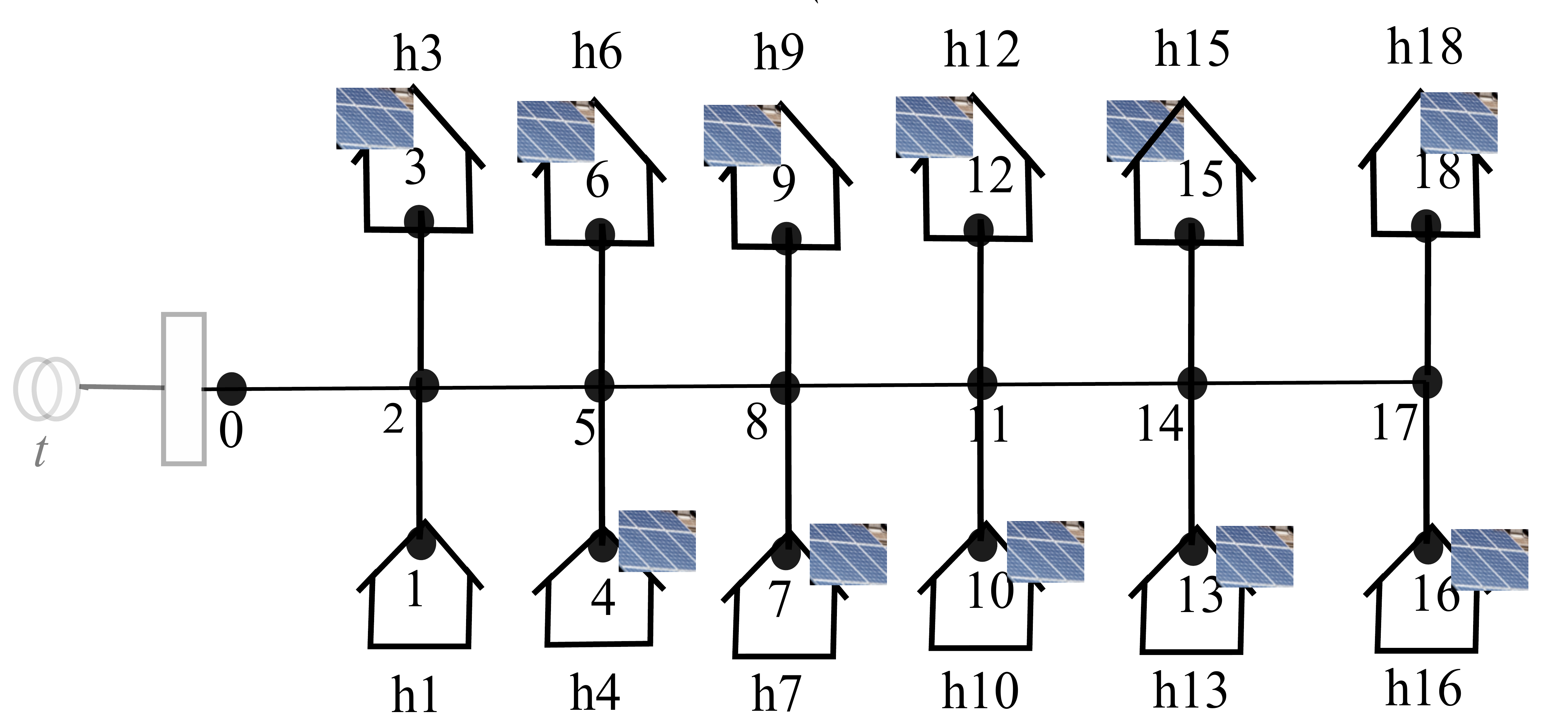}
    \vspace{-2pt}
    \caption{\small{Simplified network for numerical results}}
    \label{fig:test_grid}
\end{figure}

\begin{figure}[!htbp]
	\center
	\includegraphics[width=0.75\linewidth]{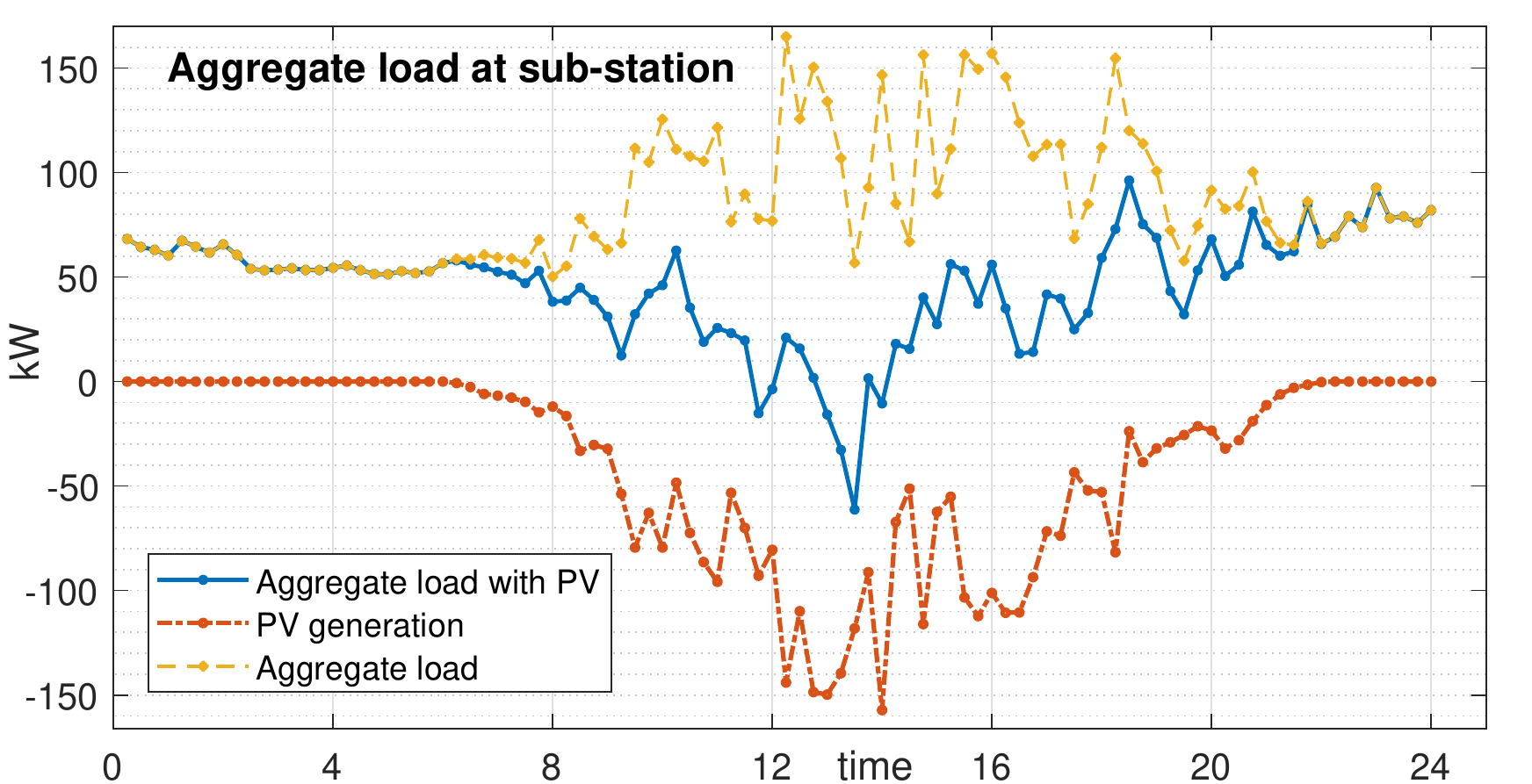}
	\vspace{-5pt}
	\caption{\small{Aggregate load, distributed PV generation as seen from substation transformer. 
	{Sampling time, $t_{\text{samp}}$, of 15 minutes.}}}
	\label{fig:aggregate}
\end{figure}

\textit{Network details:}
\label{appenix:network}
A test grid with 12 LV consumers (Fig.~\ref{fig:test_grid}) is used in the numerical illustrations. The main branch (0-2-5-8-11-14-17) is assumed to be a 150 sq mm Al cable ~300\,m per segment. The remaining branches connecting the main branch to LV consumers are assumed to be 35~sq mm Al cables of length ~150\,m each. The consumer features were selected to represent typical consumers in Belgium (Table~\ref{tab:test_LV}). The load profiles were selected from the real consumers of Belgium for a typical spring day and the PV profile was obtained from the irradiance data of the EnergyVille rooftop in Genk, Belgium.
\begin{table}[!htbp]
\caption{LV consumer in the test feeder}
\centering
\begin{tabular}{ccccc}
\hline
     prosumer& PV [kWp] & HP [kW] & SMEs & peak Load [kW]     \\ \hline
     h1 & -  & - & N & 20 \\
     h3 & 10 & - & Y & 7 \\
     h4 & 20 & - & N & 4 \\
     h6 & 8 & - & N & 2 \\
     h7 & 20 & - & Y & 9 \\
     h9 & 12 & - & N & 12 \\
     h10 & 15 & 6 & Y & 14 \\
     h12 & 12 & - & N & 14 \\
     h13 & 10 & - & N & 14 \\
     h15 & 18 & - & N& 16\\
     h16 & 18 & - & N & 20 \\
     h18 & 18 & 7.5 & N & 10 \\ \hline
\end{tabular}
    \label{tab:test_LV}
\end{table}
The aggregate load seen from the substation is shown in Fig.~\ref{fig:aggregate}. 
The mean R/X ratio of the DN equals 2.01.

\begin{figure}[!htbp]
	\center
	\includegraphics[width=0.8\linewidth]{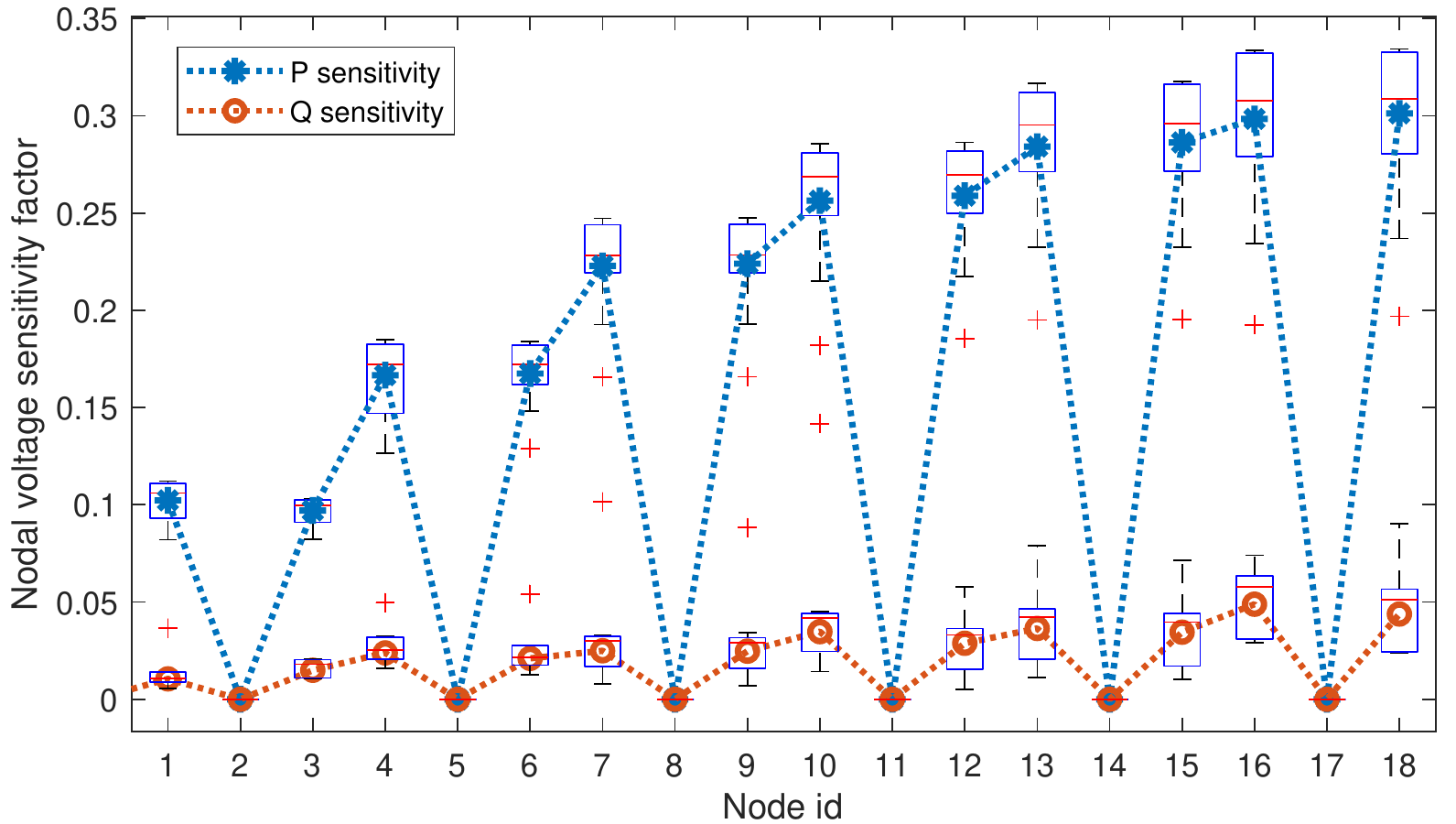}
	\vspace{-3pt}
	\caption{\small{\textcolor{black}{Active and reactive power nodal sensitivity.}}}
	\label{fig:sensitivity}
\end{figure}

\textcolor{black}{
The NVS for P and Q for the test network shown in Fig. \ref{fig:test_grid} is shown in Fig.~\ref{fig:sensitivity}.
\textcolor{black}{Note the NVS will be governed by load levels as shown in the box plots in Fig.~\ref{fig:sensitivity}. For the NVS calculation according to \eqref{eq:sensitivityMatP}, the load variations of the order of three decades are performed in order to capture the NVS variation at different load levels. In this work, we assume a mean value for NVS for active and reactive power fluctuations, shown in dotted blue and red lines respectively in Fig.~\ref{fig:sensitivity}.}
Note, the prosumers connected at the beginning of the feeder have a lower P and Q NVS compared to prosumers at the end of the DN.
We assume that nodes in no prosumers connected do have any load flexibility, therefore, have zero NVS.}

\textcolor{black}{The voltage limits are assumed to be $V_{\min}=0.92$ and $V_{\max} =1.08$ in per unit (p.u.). EN 50160 recommends a limit within $\pm10\%$ \cite{5625472, pqfact}. The line loading limit is assumed to be 100\% of the rated capacity of the line. The permissible limits are $\Delta V_{\text{perm}} = 0.04$ p.u. and $\Delta T_{\text{perm}} = 75\%$. The tuning of these levels are beyond the scope of this paper and needs further analysis.}

\textit{OPF duals vs flexibility activation signals:}

\begin{figure}[!htbp]
	\center
	\includegraphics[width=0.76\linewidth]{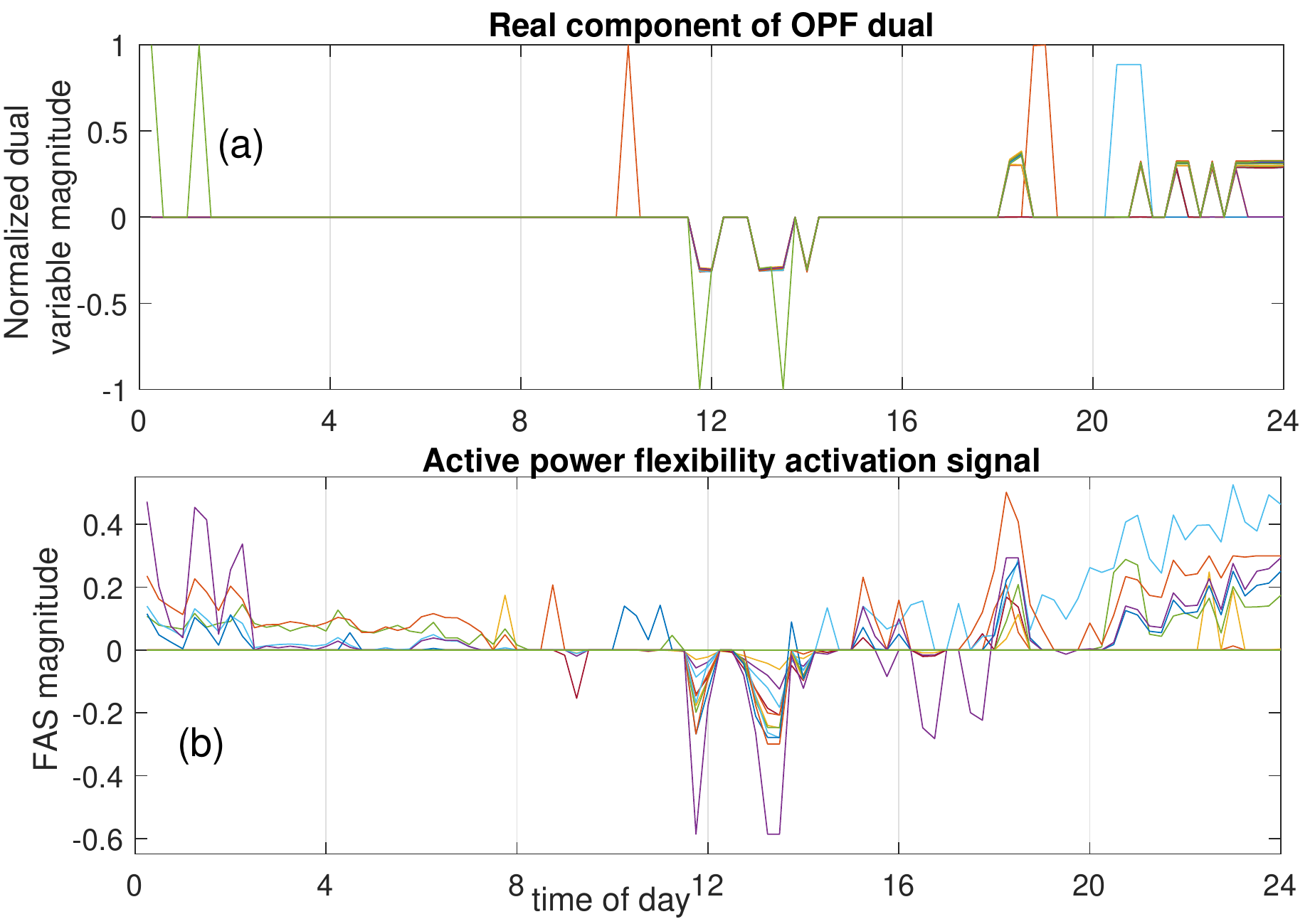}
	\vspace{-7pt}
	\caption{\small{OPF duals vs FAS. 
	\textcolor{black}{The first plot (a) shows the OPF dual associated with constraint \eqref{eq:powerbalance} }
	The second plot (b) shows the nodal flexibility value plotted over an entire day. 
	{Each color in (a) and (b) denotes \textcolor{black}{a} dual variable and FAS for a particular node of the network.}}}
	\label{fig:dualvsflex}
\end{figure}

The dual variables of the optimal power flow problem are often used as locational marginal prices (LMP).
\textcolor{black}{
Our proposed flexibility activation signals hold some similarities, as can be observed in Fig.~\ref{fig:dualvsflex}.
LMPs are traditionally defined as the dual variables corresponding to the power balance equation \cite{1198282}.
\textcolor{black}{
Since the RDOPF formulation does not consider unit commitment constraints such as generator starting time, and ramping constraints of the generator, therefore, load and generation variations are instantly met with a change in substation power levels provided the aggregate power output of the transformer lies within the range of maximum and minimum limits. Thus, the dual variables associated with \eqref{eq:powerbalance} for RDOPF are  active during voltage and thermal violations, as the FAS design incentivizes resource activation during those periods. Note that the RDOPF formulation includes the penalty associated with losses in the DN, which is sensitive to the changes in the demand. In order to eliminate this impact, the dual variable is calculated using standard full AC OPF to compare with FAS generated, shown in Fig.~\ref{fig:dualvsflex}.}}
The dual variables, \textcolor{black}{shown in Fig.~\ref{fig:dualvsflex}(a), do not provide corrective feedback prior to the need for resource activation, which is reflected as infeasible OPF.
Proposed flexibility activation signals unlike the OPF duals actively try to correct network flow and voltage levels if they exceed permissible safe levels of operation.}

Next, we detail three case studies using RDOPF.
\vspace{-5pt}
\subsection{\textbf{Case study 1}: Shortcomings of SOC relaxed RDOPF formulation and tuning {line losses penalty}}
The AC formulation for RDOPF is a nonlinear optimization problem.
{Due to SOC approximation, the voltage phase information is lost. Several prior works have concluded the SOC approximation could lead to artificially inflating line losses \cite{zolfaghari2018bilevel, zhang2017novel}. These observations are almost entirely observed for traditional OPF problem formulations, where DN load is considered a parameter and not a state variable. In {the} presence of load or generation flexibilities, the DN load could vary. In such a case, SOC approximation will disproportionately activate load curtailment compared to generation curtailment. To avoid such a condition, RDOPF considers penalizing DN losses. However, a high loss penalty would lead to {increased} 
activation of DN flexible resources. This case study deals with a mechanism to tune {the loss penalty for RDOPF to reduce the optimality gap} for the AC and SOC formulation {of RDOPF} while reducing the amount of {flexibility} needed.
Next, we describe the performance indices used in this case study.}\\

\subsubsection{Performance indices}

{To} analyse the numerical simulations, we use the following performance indices:
\begin{itemize}
    \item \textit{Objective function value}: denotes the value of the cost function used in optimizing resource activation. It is denoted as $\texttt{Obj}_{\text{ACP}}$ and $\texttt{Obj}_{\text{SOC}}$ for AC formulation of RDOPF and SOC relaxed RDOPF formulation, respectively.
    \item \textit{Optimality gap} for SOC {formulation is given as}
    \begin{equation}
        \text{Optimality gap in \%} = \frac{\texttt{Obj}_{\text{ACP}} - \texttt{Obj}_{\text{SOC}}}{\texttt{Obj}_{\text{ACP}}} \times 100.
    \end{equation}
    \item \textit{Branch losses}: SOC relaxed RDOPF could lead to fictitious line losses{, and this is because of the conic voltage limit}. The {resultant} nodal voltage is slightly lower than the one from {AC RDOPF}. {To} compare, we define:
    \begin{itemize}
        \item Cumulative line loss in all branches overall time is denoted as 
        \begin{equation}
            \text{Loss}_{\text{cumulative}} = \sum_t \sum_{i,j} L_{i,j,t},
        \end{equation}
        \item Mean voltage deviation in percentage is given as
        \begin{equation}
            \Delta V^{\text{SOC}} = \frac{ \sum_t \sum_i \Delta V_{i,t}^{\text{SOC}}}{\text{number of samples}}.
        \end{equation}
    \end{itemize}
    \item Cost of losses is denoted as
    \begin{equation}
        \text{Cost}_{\text{loss}} = \lambda_{\text{loss}}\sum_t  \rho(t),
    \end{equation}
where $\lambda_{\text{loss}}$ denotes line losses penalty.
\end{itemize}

\subsubsection{Numerical result}
\begin{figure}[!htbp]
	\center
	\includegraphics[width=0.85\linewidth]{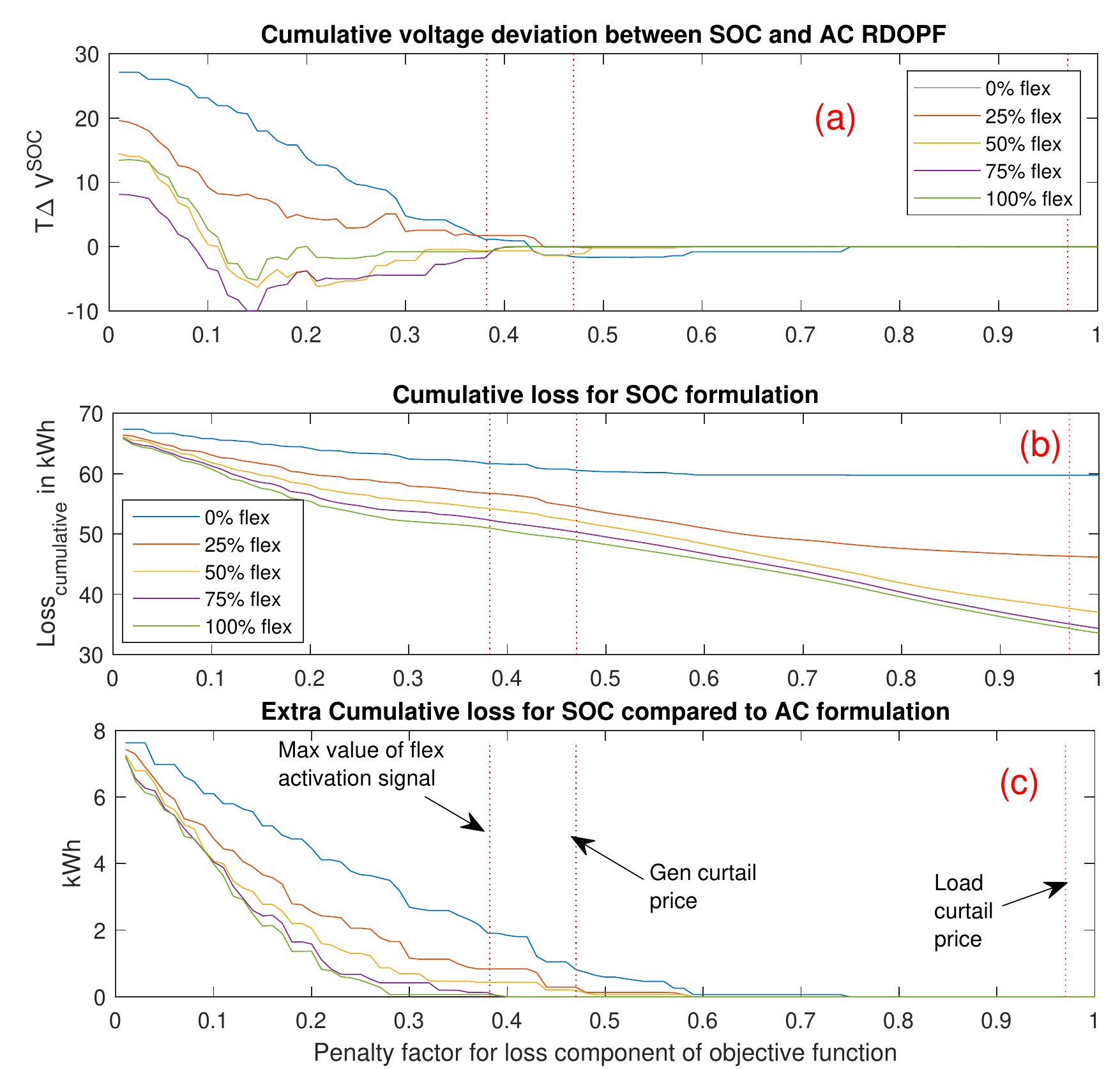}
	\vspace{-5pt}
	\caption{\small{Observe the DN losses decreases as the penalty factor increases.
	(a) cumulative voltage deviation for SOC formulation when compared with {AC} RDOPF.
	(b) the cumulative line losses in kWh for different flexibility levels. (c) the {extra loss in SOC compared to AC RDOPF.} }}
	\label{fig:soclosses}
\end{figure}

In some cases,
SOC approximations in flexibility activation could cause numerical errors by increasing artificial losses in branch flows. 
Figure~\ref{fig:soclosses} shows {how the loss penalty component in the objective function influences the} (a) cumulative voltage deviation, (b) cumulative losses for SOC {relaxed} RDOPF, and (c) extra losses for SOC RDOPF compared to the {AC} RDOPF. 
It can be observed that for a loss penalty greater than 0.6, the gap between SOC and AC formulation of RDOPF is negligible. 
Note the generator and load curtailment costs of 0.47 and 0.87, respectively.


\begin{figure*}[!htbp]
	\center
	\includegraphics[width=0.952\linewidth]{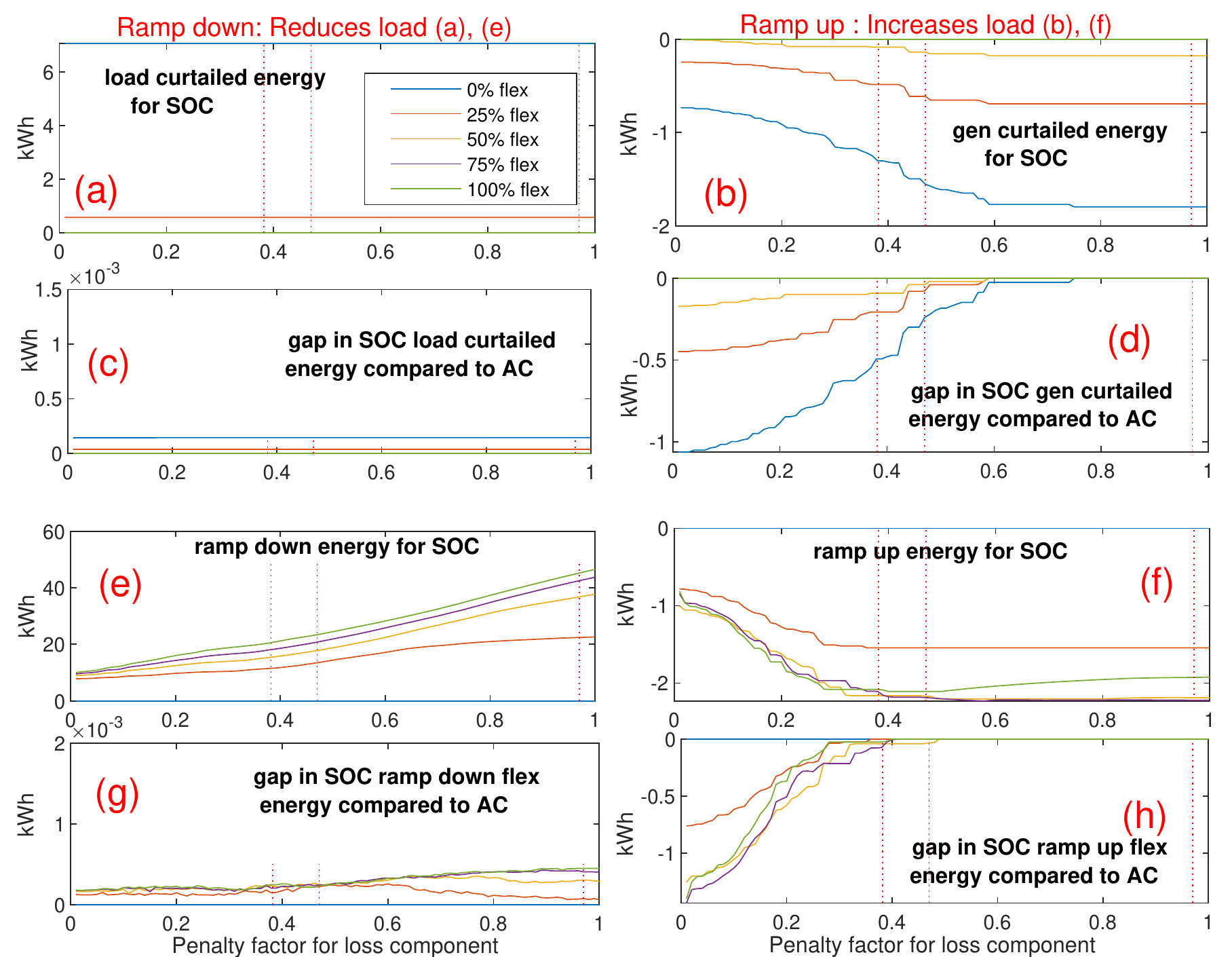}
	\caption{\small{The comparison of resource activation for SOC and RDOPF formulations \textcolor{black}{for varying loss penalty factor. (a) shows the load-curtailed energy, (b) shows the generation-curtailed energy for SOC formulation, (c) and (d) show the gap in load and generation-curtailed energy gap between SOC and AC RDOPF, (e) and (f) show the ramp-down and up energy and (g) and (h) shows the gap between SOC and AC RDOPF. }
	}}
	\label{fig:socresourceact}
\end{figure*}

Fig.~\ref{fig:socresourceact} shows the ramp up and ramp down resource activation for SOC formulation and the difference {from} the AC RDOPF.
Fig.~\ref{fig:socresourceact}~(c) and (g) denotes that irrespective of loss penalty, the {required load curtailment and ramp-down energy obtained from} SOC RDOPF matches that from AC RDOPF. 
However, SOC formulation leads to a reduced amount of generation curtailment and reduced activation of ramp {up} resources, which reduces nodal load, 
 see Fig. \ref{fig:socresourceact} (d) and (h). 
 {Note from Fig. \ref{fig:socresourceact} (e), as the penalty factor increases, a greater amount of ramp-down flexible resources are activated, leading to a reduction in nodal load, thus reducing flow losses. A similar trend is not observed for Fig. \ref{fig:socresourceact} (f) and (b). This implies SOC formulation is trying to reduce the nodal load by activating more amount of flexibility as the cost of losses increases.}
This is caused due to inconsistent voltage differences between nodes connecting a branch in the case of the SOC formulation compared to AC {RDOPF}.
The aberration in branch voltage difference causes fictitious losses in the DN, which causes the optimality gap for SOC flex formulation to increase. 
This gap can be reduced by increasing the loss penalty.
However, this fix leads to increased flexibility activation of the type, which reduces the net load of the DN. 

\subsubsection{Fixing loss penalty factor}
Previously we observed that increasing loss penalty leads to:
\begin{itemize}
    \item Reduction in optimality gap between SOC and AC RDOPF: {observe in  Fig. \ref{fig:socresourceact} (c), (d), (g) and (h) converges to zero as $\lambda_{\text{loss}}$ increases. This implies, with high enough $\lambda_{\text{loss}}$, the SOC and AC RDOPF flexibility outputs converge.}
    \item Decreased activation of generation curtailment and ramping up flexible resources and increased activation of ramping down 
    (reduce nodal load)
    flexible resources, see Fig.~\ref{fig:socresourceact}.
\end{itemize}

\begin{figure}[!htbp]
	\center
	\includegraphics[width=0.92\linewidth]{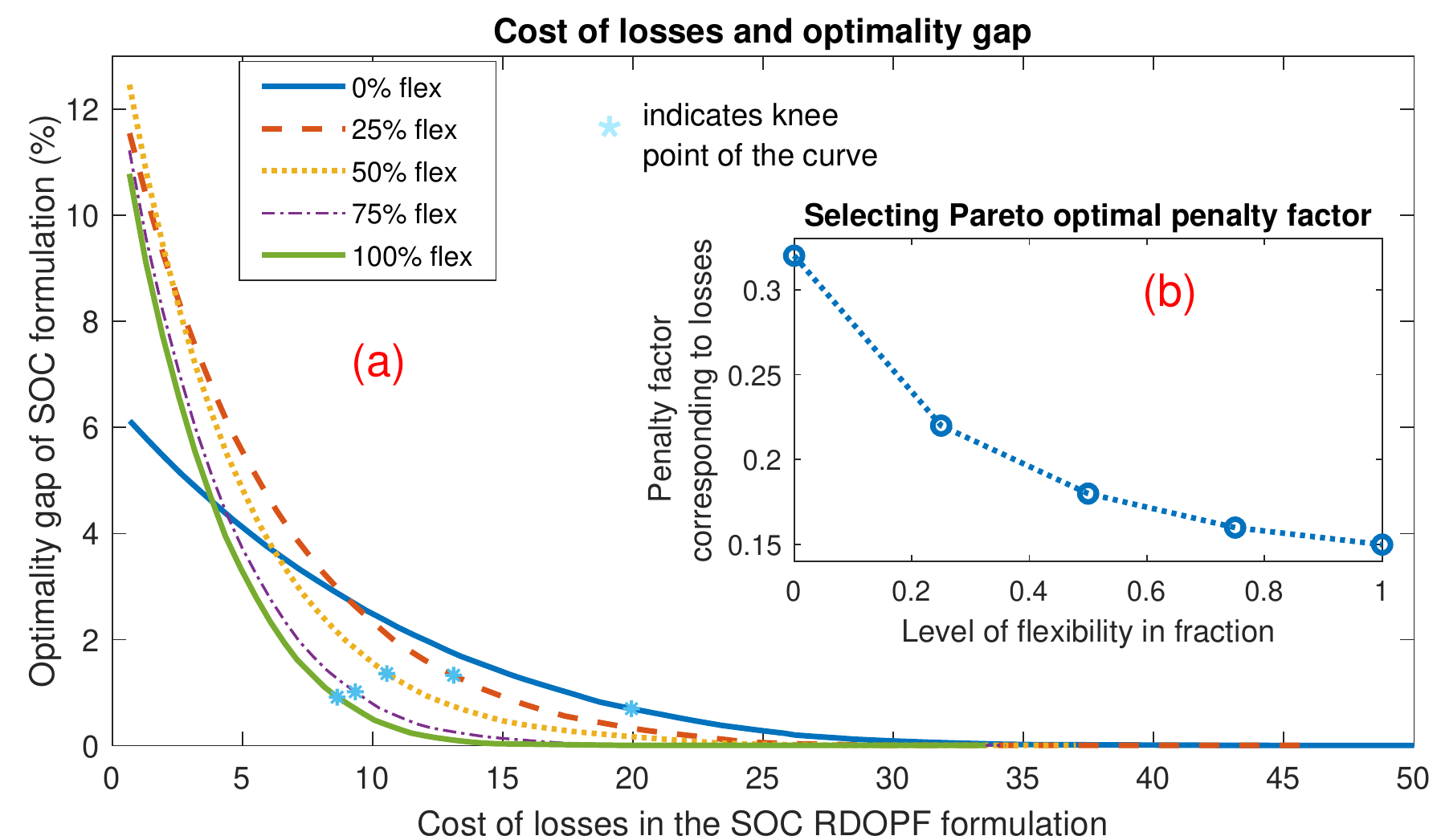}
	\vspace{-5pt}
	\caption{\small{Fixing penalty factor using Pareto optimality. (a) shows the indifference curve with optimality gap on y-axis and $\text{Cost}_{\text{loss}}$ on x-axis. (b) Knee point is calculated from the plot (a) and corresponding $\lambda_{\text{loss}}$ values are plotted.}
	}
	\label{fig:socpareto}
\end{figure}
\begin{figure*}[!htbp]
	\center
	\includegraphics[width=0.4\linewidth, angle=90]{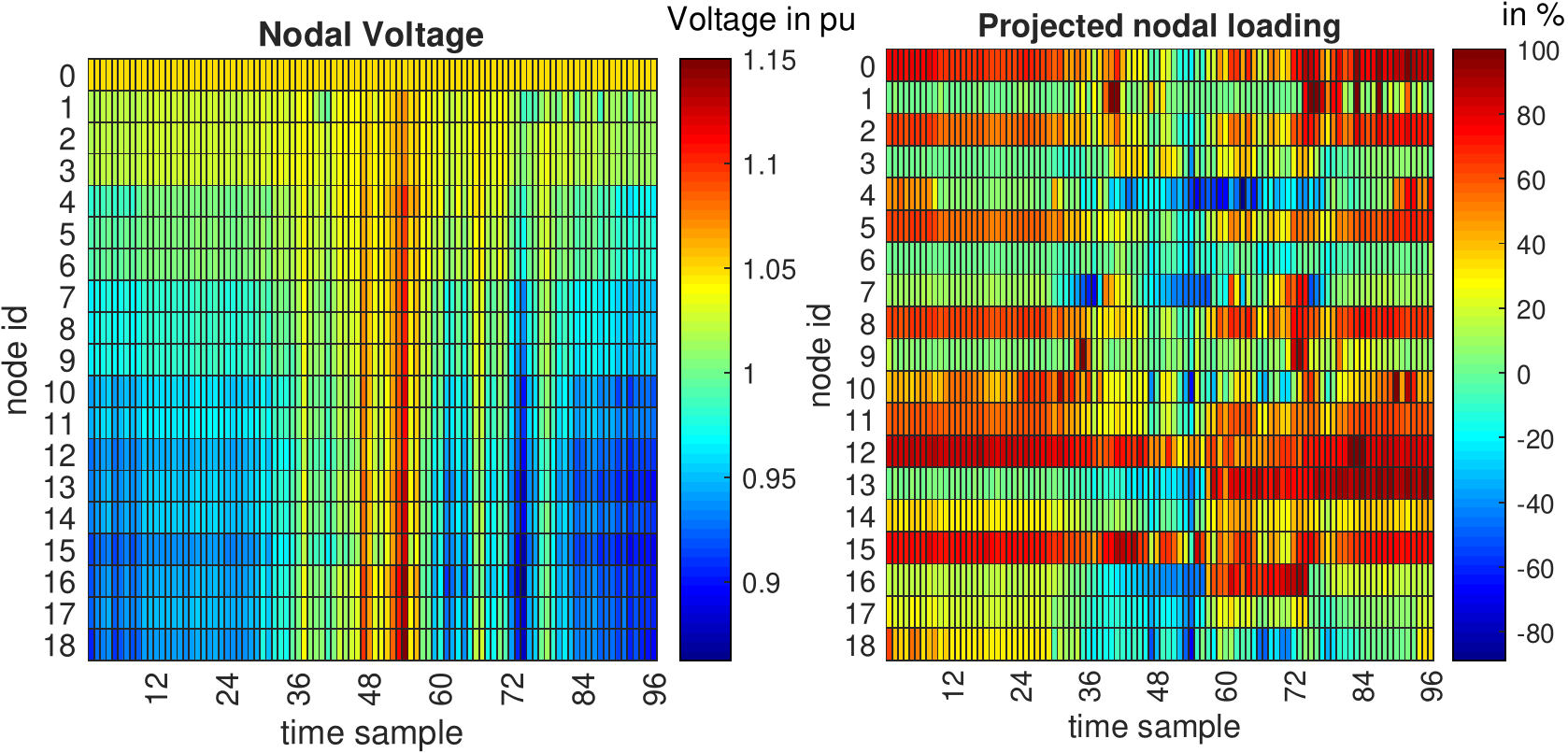}
	\caption{\small{DN state in the absence of load flexibility and curtailable resources.}}
	\label{fig:c2network}
\end{figure*}

\textcolor{black}{
In multi-objective optimizations where some objectives are in conflict with others, that is, increasing or decreasing some objective components adversely impacts some other objective components, then in such a case Pareto optimality can be used to identify suboptimal individual objective components which are collectively optimal \cite{1599245}.
Prior works such as \cite{knezovic2017robust} utilize a Pareto front for an electric vehicle charging problem for an aggregator while considering DSO's goal towards reducing the charging cost and losses.}
The goal here is to identify the Pareto optimality loss penalty value which (a) reduces the optimality gap while (b) reduces the cost of losses.
\textcolor{black}{The Pareto surface shown in Fig. \ref{fig:socpareto} is identified 
by performing a sensitivity analysis by varying loss penalty magnitude for different flexibility levels.}
As observed in Fig. \ref{fig:socpareto}, the two goals of reducing the optimality gap and minimizing loss cost are competing with each other, implying if we improve one of them, we sacrifice the other. Fig. \ref{fig:socpareto}(a) shows the value of the optimality gap and cost of losses in SOC formulation. It can be observed that the optimality gap increases beyond 12\% for low-loss costs. However, for a very high loss cost, the optimality gap converges to zero.
The knee point of the plot, \textcolor{black}{calculated using \cite{kneepoint}}, is shown in Figure \ref{fig:socpareto}(a) and is considered as the best-suited value of the loss penalty. In Figure \ref{fig:socpareto}(b) the optimal loss penalty values are plotted for different levels of flexibility. It is observed that the loss penalty value deteriorates with an increase in the level of flexibility.

\subsection{\textbf{Case study 2}: Temporal and locational flexibility and curtailment needs assessment}
From a DSO perspective, knowing the location and temporal attributes of flexibility or curtailment needs is crucial \textcolor{black}{to take effective grid management}. It will assist the DSO in planning \textcolor{black}{flexible resources} and procuring the best resources in energy markets, and thus avoid unforeseen network events in due time.
This numerical case study explores activated resources' temporal and locational attributes: flexible and curtailable. 
{Case study 2 and 3 use AC RDOPF.}
The feasibility of AC RDOPF is verified using \textcolor{black}{full AC} power flows for this numerical case.

\subsubsection{Metrics for DN resource needs assessment}
The needs assessment of flexible and curtailable resources in DN planning and operation is critical for its reliable operation.
The flexible and curtailable resource needs assessment is performed based on three metrics: (a) the aggregate metrics, (b) locational metrics, and (c) temporal metrics.\\

\textit{The aggregate flexibility and curtailment metrics}
detail the cumulative energy and maximum power levels of flexibility and curtailment {across all times and nodes of DN}. The following indices are used:
\begin{itemize}
\item Metrics for generation and load curtailment:
(a) Cumulative generation curtailed energy,
(b) Cumulative load curtailed energy,
(c) Maximum generation curtailment power,
(d) Maximum load curtailment power.
    \item Metrics for ramp-up and down flexibility:
(i) Cumulative ramp-up flexibility energy,
(ii) Cumulative ramp-down flexibility energy,
(iii) Maximum ramp-up flexibility power,
(iv) Maximum ramp-down flexibility power.
\end{itemize}
These aggregate metrics show the overall need for flexible or curtailable resources in a DN.
However, aggregate metrics do not provide any locational or temporal resource needs of the distribution network.\\

\textit{Locational and temporal metrics:}
The output of RDOPF is four $N\times T$\footnote{$N$ denotes the number of nodes, i.e. cardinality of the set of nodes denotes by $\mathscr{N}$, and $T$ denotes the number of time instances. Nodes without flexibility and/or curtailment will have a value equal to zero.} order matrices: $R_{\text{up}}, R_{\text{down}}, C_{\text{load}}$ and $C_{\text{gen}}$.
These matrices denote ramp-up flexibility, ramp-down flexibility, load curtailment, and generation curtailment activated at the least possible cost for achieving a feasible solution for RDOPF.
For an arbitrary set $\mathcal{X} {=} \{R_{\text{up}}, R_{\text{down}}, C_{\text{load}}, C_{\text{gen}} \}$, the temporal and locational needs for time $i$ and node $j$ are given as:
$
    X^{\text{temporal}}_i = \sum_{y=1}^T X(y,i),~~~
    X^{\text{locational}}_j = \sum_{y=1}^N X(j,y).
$
{For example, temporal ramp up needs are given as ${R^{\text{temporal}}_{\text{up},i}}=\sum_{y=1}^T R_\text{up}(y,i)$.}\\

\begin{figure}[!htbp]
	\center
	\includegraphics[width=\linewidth]{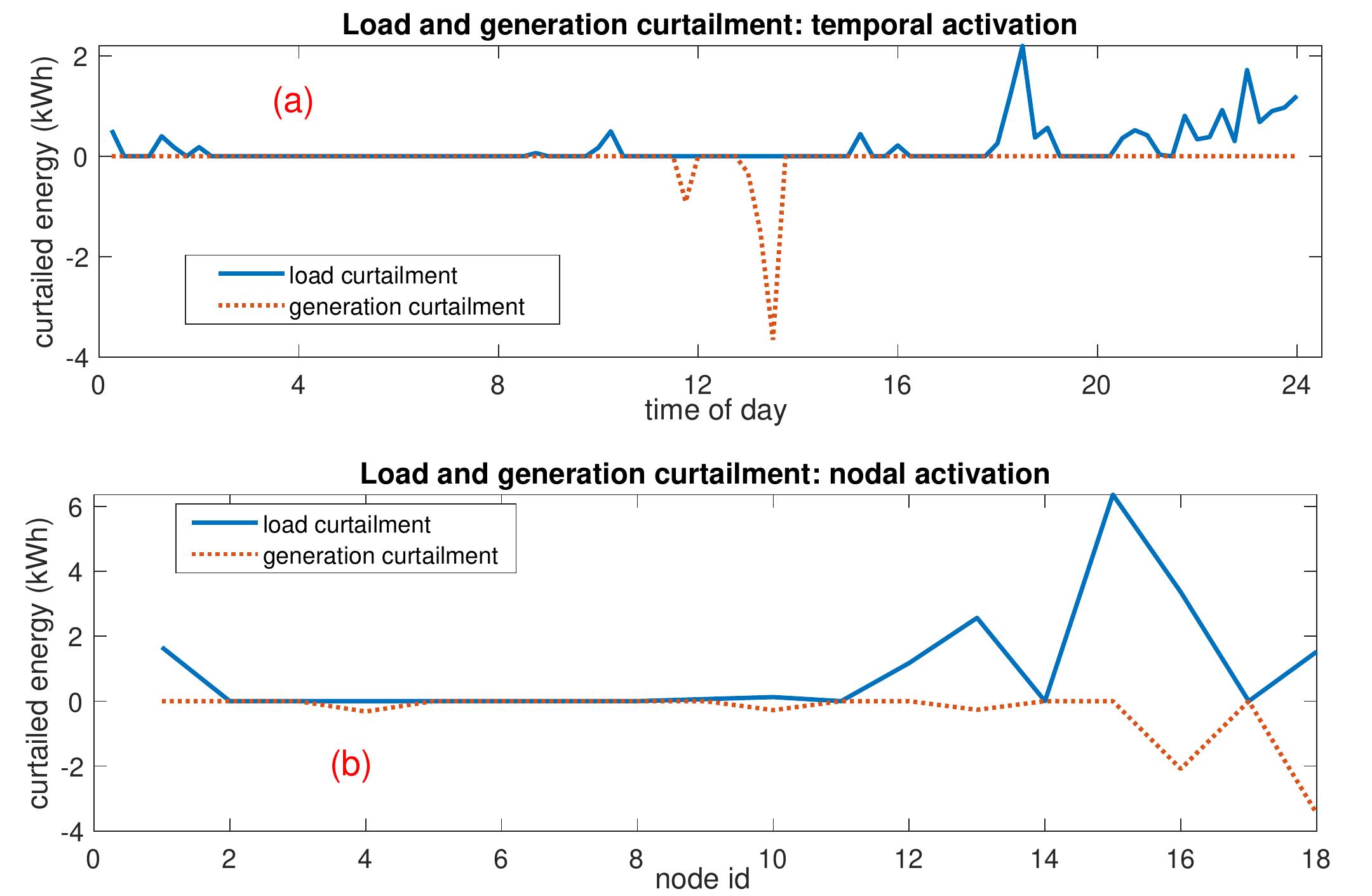}
	\vspace{-3pt}
	\caption{{Load and generation curtailment in the absence of flexibility. \textcolor{black}{Plot (a) is aggregated over all nodes, and plot (b) is aggregated over all time.}} }
	\label{fig:c2curtnoflex}
\end{figure}
\begin{figure*}[!htbp]
	\center
	\includegraphics[width=6.5in]{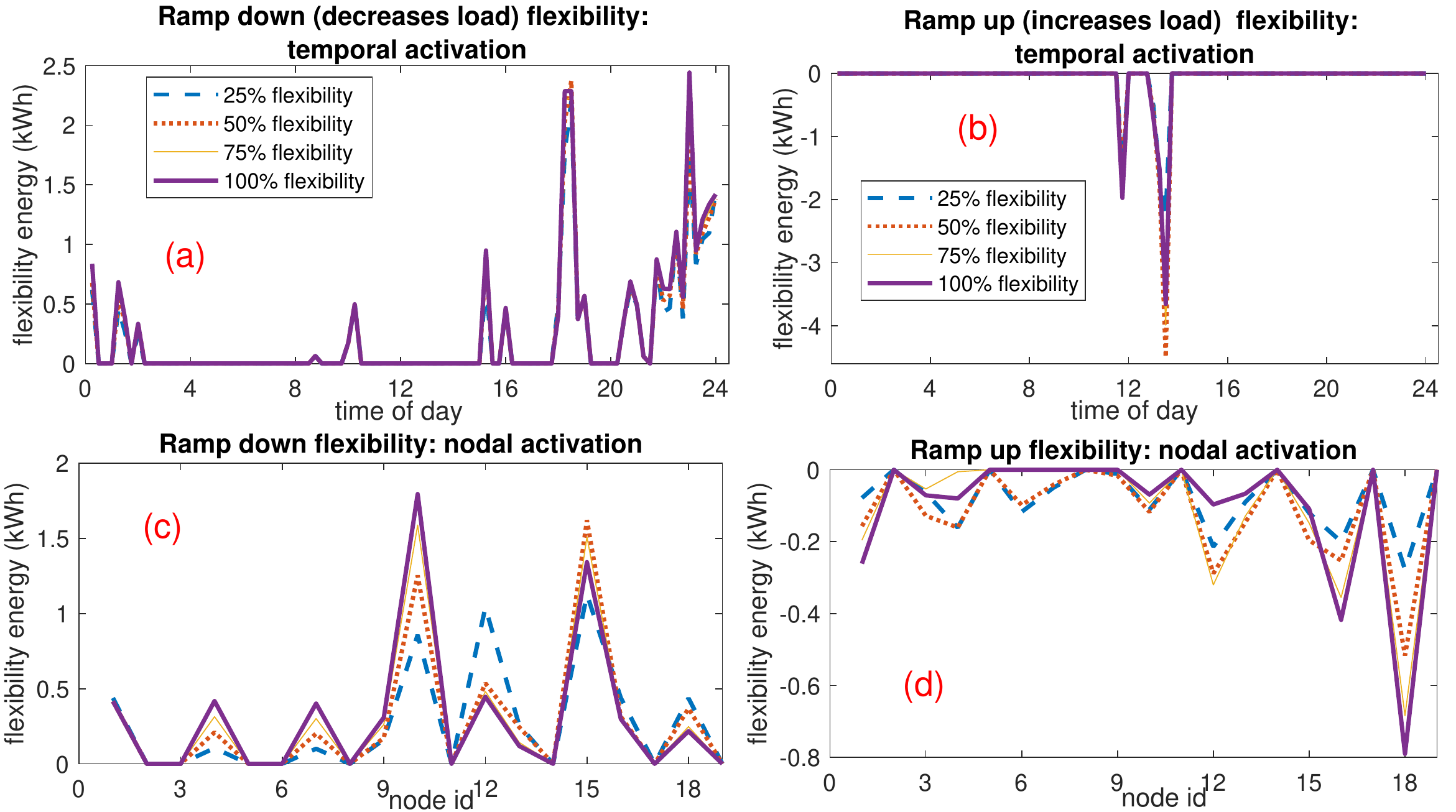}
	\vspace{-5pt}
	\caption{\small{Ramp up and ramp down flexibility, locational and temporal needs assessment with varying levels of flexibility.}}
	\label{fig:c2flexact}
\end{figure*}
\begin{figure}[!htbp]
	\center
	\includegraphics[width=0.8\linewidth]{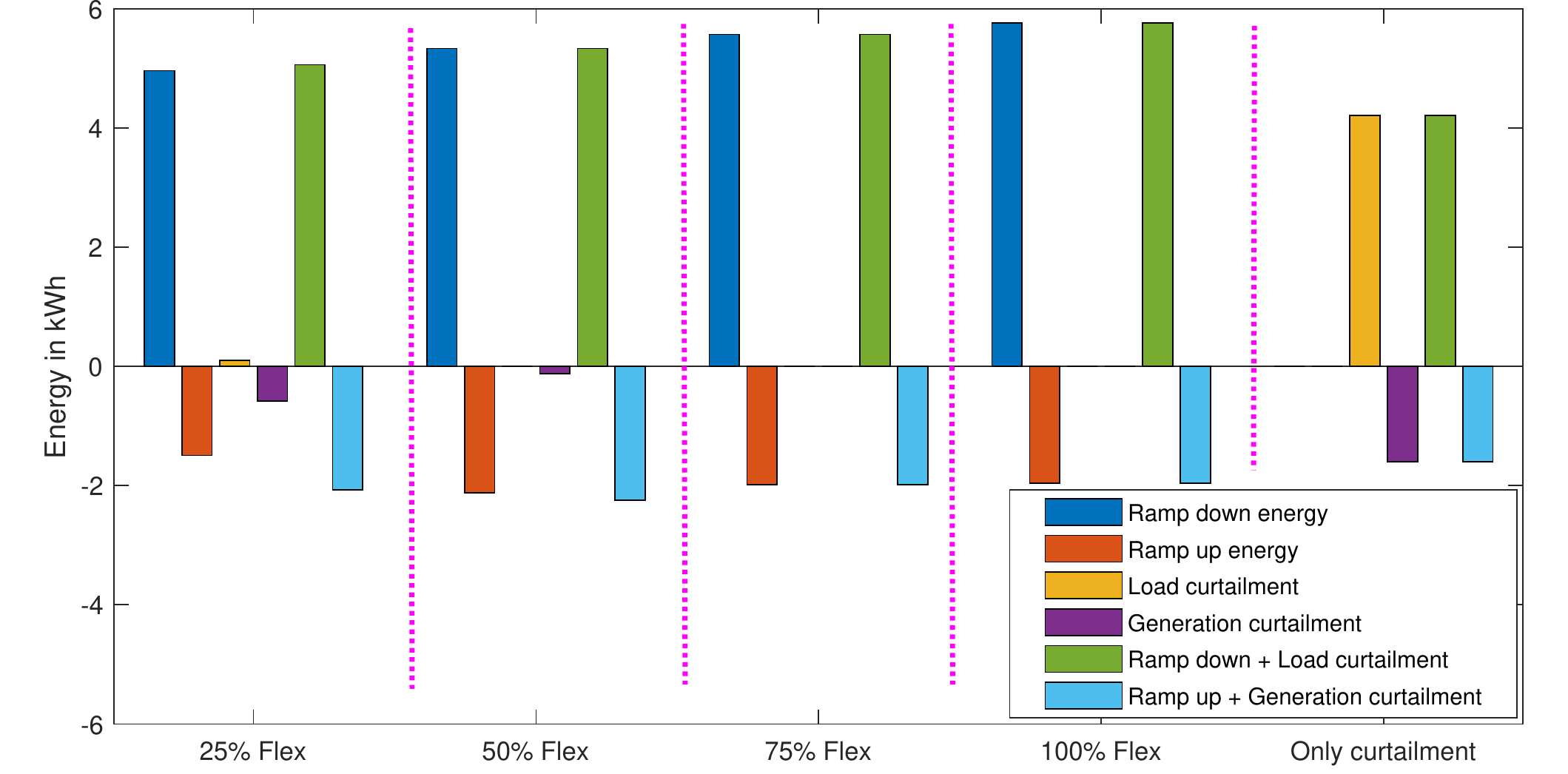}
	\vspace{-1pt}
	\caption{\small{Cumulative flexibility and curtailment activation with 25, 50, 75, 100, and 0\% flexibility levels along with load and generation curtailment.}}
	\label{fig:bar}
\end{figure}

\begin{figure*}[!htbp]
	\center
	\includegraphics[width=0.65\linewidth, angle=90]{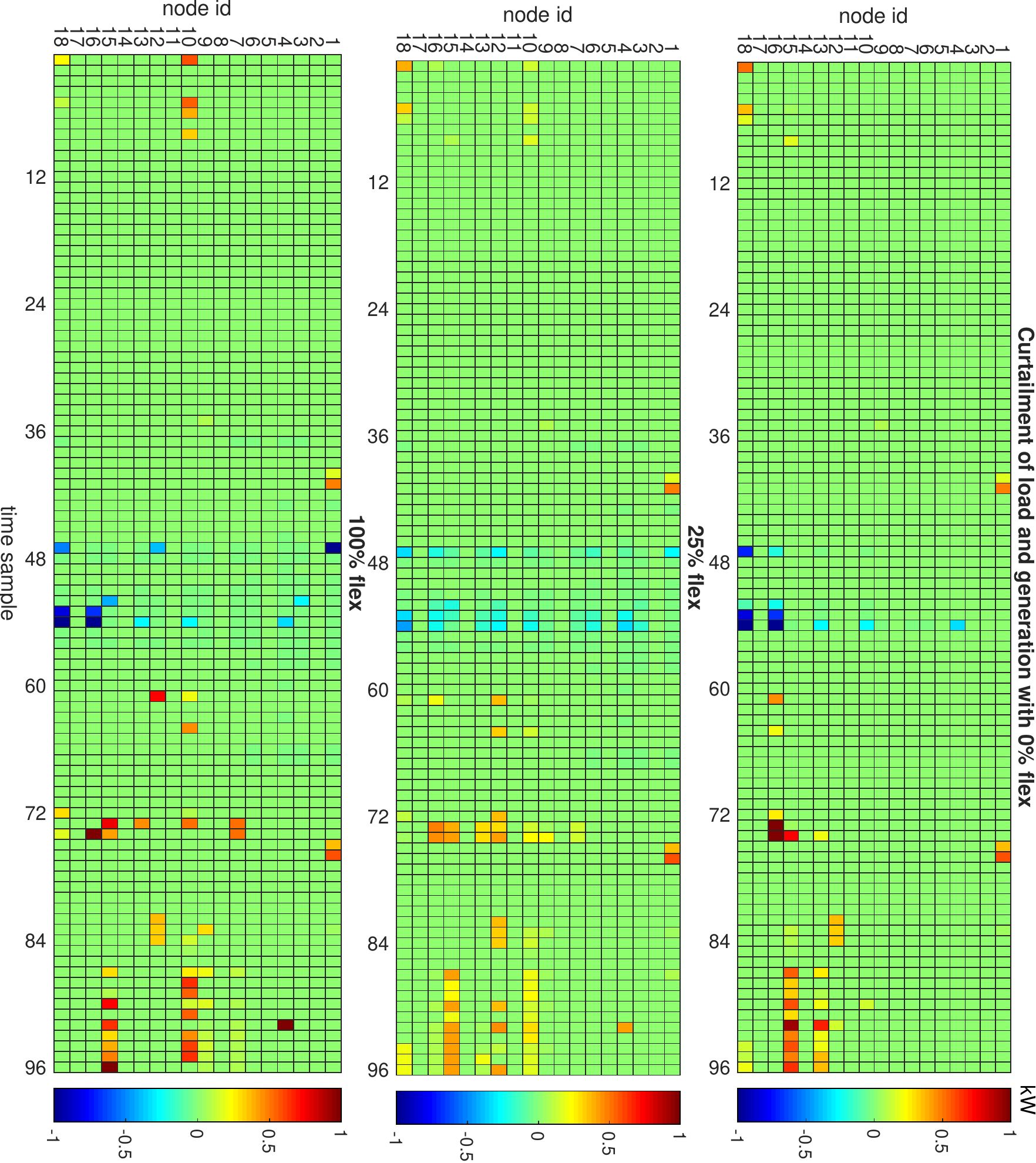}
	\vspace{-1pt}
	\caption{\small{Temporal and locational activation of curtailment and flexibility \textit{power} in kW. Heatmaps are shown in (a) curtailment of load and generation at 0\% flexibility levels, (b) flexibility activation with its level of 25\%, (c) flexibility activation with its level of 100\%.}}
	\label{fig:c2flexpower}
\end{figure*}

\subsubsection{Numerical results}

Fig.~\ref{fig:c2network} summarizes the network state using heatmaps for nodal voltages and projected nodal loadings.
In the nodal voltage plot, we can observe under-voltage cases in the early morning and the evening. 
During the day, due to solar generation, over-voltage instances can be observed.
In the projected nodal loadings heatmap, nodes 1, 12, 13, and 15 {(non-exhaustive list)} 
are overloaded for some part of the day. 
Reverse power flow \textcolor{black}{caused by PV generation injection into the DN} can be observed during the day at nodes 4, 7, 16, and 18.
Note that there are no flexible and curtailable resources connected in this nominal case, and power flow calculations are performed to understand the network state. 

Table \ref{tab:resourceactiva} summarizes the  performance indices for the nominal case with 0\%, 25\%, 50\%, 75\%, and 100\% flexibility for all nodes along with load and generation curtailment. 
Note that in these numerical results, the curtailed power {is} bounded by only the load magnitude at the node.
For the nominal case, 2.02\% of the time over-voltage is observed, 5.37\% of the time under-voltage is observed, and 0.75\% of the time the branches have a loading exceeding or equal to 100\% of the line rating.
With flexible and curtailable resource activation, these network issues are resolved as listed in Table \ref{tab:resourceactiva}. 
Table \ref{tab:resourceactiva} summarizes the cumulative curtailment and flexibility requirements of the DN. 
Note that with the increase in flexible resources, the objective function value decreases, implying activation cost reduction.
Further, the cumulative line losses slightly reduce from 13.17 kWh to 13.01 kWh.
Figure \ref{fig:c2curtnoflex} summarizes the cumulative curtailable resource needs of the distribution network in terms of the time of day and location {in} the network.
These curtailable resource requirements are calculated in the absence of ramping up and ramping down flexibility \textcolor{black}{limits}.
Figure \ref{fig:c2flexact} presents the ramp-up and ramp-down activated flexibility in terms of energy quantified based on the time and location of activation.
It can be observed from Figure \ref{fig:c2curtnoflex} and Figure \ref{fig:c2flexact} that the flexibility needs can be approximated based on the unbounded curtailment needs of the network. 

\begin{center}
\begin{table*}[htb]
\scriptsize
\centering
\small
	\caption {Performance indices for without and with flexible and curtailable resource activation} 
	\label{tab:resourceactiva}
	\begin{center}
		\begin{tabular}{lcccccc}
			\hline
			
			\multirow{2}{0.3\columnwidth}{Performance index}& Without flex, curt & \multicolumn{5}{c}{ {With load and generation curtailment}} \\
			 & Nominal Case & 0\% flex & 25\% flex & 50\% flex & 75\% flex & 100\% flex\\
			\hline
			\hline
			Load curtailed energy (kWh) &  -  &  4.21   &    0.09   &   0    &   0    &     0     \\
			Generation curtailed energy (kWh) & - & -1.60    &    -0.58   &   -0.12    &    0   &    0      \\
			Nodal load curtailed peak power (kW) & - &   1.27    &    0.39   &   0    &    0   & 0        \\
			Generation curtailed peak power (kW) &  -     &   -1.80    &  -1.21     &    -0.43   &    0   &  0 \\
			\hline
			Flex ramp down energy (kWh) &   -    &   0 & 4.96   &   5.34    &   5.57    &      5.76   \\
			Flex ramp up energy (kWh) &    -   &  0     &   -1.50    &   -2.12    &    -1.99   &  -1.96 \\
			Flex ramp down peak power (kW) &  -     &    0   &   0.57    &   0.92    &    1.26   &  1.67 \\
			Flex ramp up peak power (kW) &    -   &   0    &  -0.46     &   -0.92    &    -1.38   &  -1.78 \\
			\hline
			Samples voltage $>V_{\max}$ (\%)  &   2.02    &    0   &     0  &   0    &     0  & 0   \\
			Samples voltage $<V_{\min}$ (\%)  &   5.37    &     0  &   0    &   0    &    0   & 0   \\
			Samples voltage $>1+\Delta_{\text{perm}}$ (\%)  &   14.96    &   14.97    &  14.91     &   14.91    &  14.91     &  14.91 \\
			Samples voltage $<1-\Delta_{\text{perm}}$ (\%)  &   28.67    &  26.64     &    26.48   &    26.48   &    26.48   &  26.48 \\
			Samples thermal loading $\geq 100$ (\%)  &   0.75    &   0    &    0   &   0    &    0   &   0\\
			Samples thermal loading $>75$ (\%)  &   13.48    &    9.05   &   8.94    &   9.21    &   9.21    &  9.16 \\
			Cumulative line losses (kWh)  &    -   &    13.17   &   13.05    &    13.02   &   13.02    &  13.01 \\
			\hline
			Cost of load curtailment (\euro) &   -    &    16.33   &  0.38     &    0   &    0   &  0 \\
			Cost of ramp down activation (\euro) &   -    &   0    &    3.74   &   3.49    &  3.355     & 3.259  \\
			Cost of generation curtailment (\euro) &   -    &   3.01    &  1.09     &   0.23    & 0      & 0  \\
			Cost of ramp up activation (\euro) &  -     &    0   &   0.99    &  1.42     &  1.43     &  1.35 \\
			\hline
			Solve time (seconds) &   -    &   4.29    &   5.36    &    5.58   &   5.13    & 4.97   \\
			Objective function value &   -    &   19.35    &  6.19     &   5.14    &   4.78    &   4.61\\
             \hline
		\end{tabular}
		\hfill\
	\end{center}
\end{table*}
\end{center}

Fig.~\ref{fig:c2flexact} shows the ramp-up and ramp-down temporal and locational flexibility activation with varying levels of flexibility.
Although the temporal activation plot for Fig.~\ref{fig:c2flexact} (a) and (b) resembles that of Fig.~\ref{fig:c2curtnoflex} (a). However, location activation of flexibility shown in Fig.~\ref{fig:c2flexact} (c) and (d) differs from that of Fig.~\ref{fig:c2curtnoflex} (b). The curtailment needs, and flexibility needs differ because FAS is varying based on location and time as it is governed by NVS and network state. However, the curtailment cost is flat. Therefore, curtailment needs of the DN will form the lower limit of flexibility needs of the DN, see Fig.~\ref{fig:bar}.

\begin{figure}[!htbp]
	\center
	\includegraphics[width=0.8\linewidth]{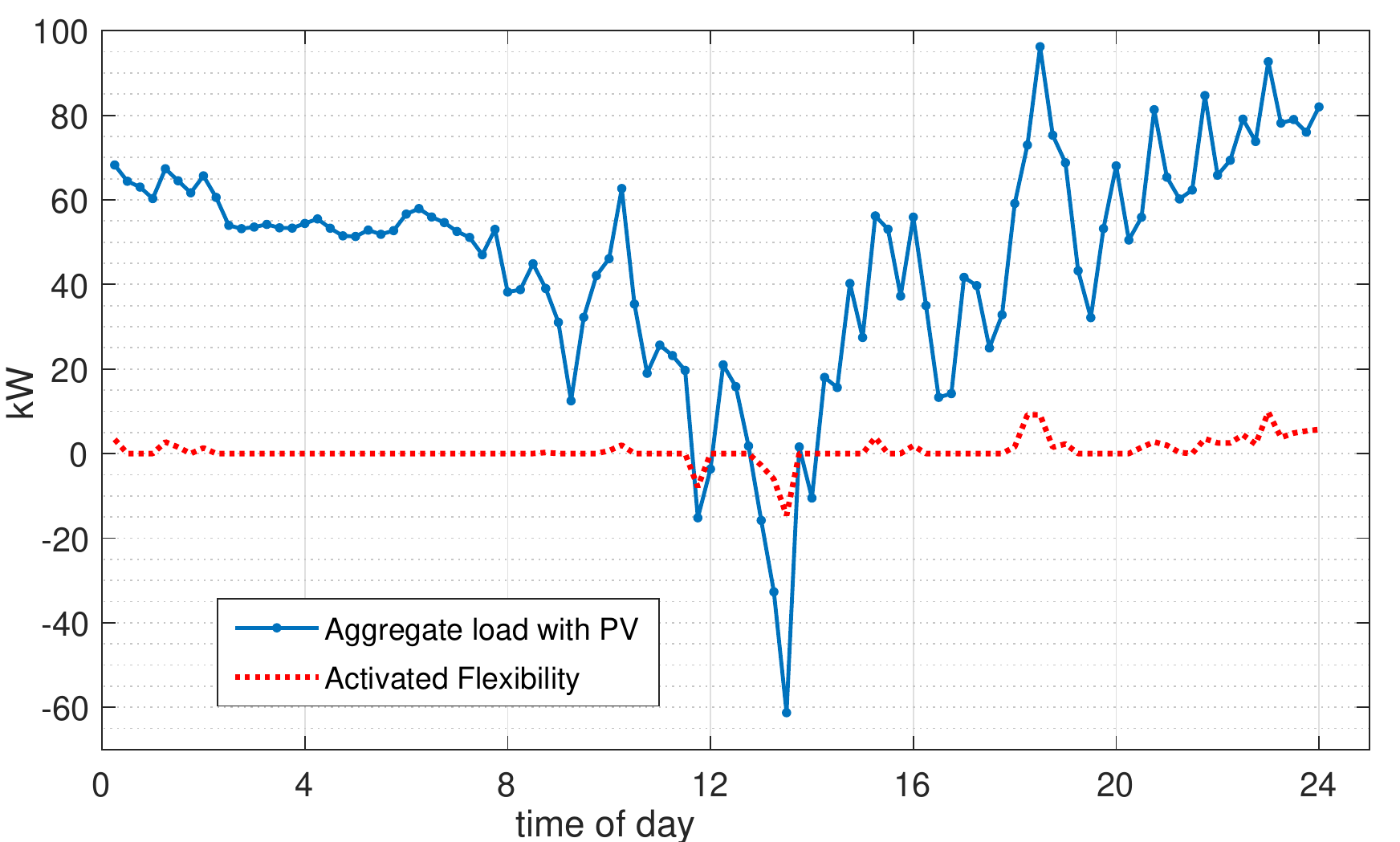}
	\vspace{-4pt}
	\caption{\small{\textcolor{black}{The activated flexibility when compared with the aggregated load served by the sub-station.}}}
	\label{fig:flexact}
\end{figure}

Figure \ref{fig:c2flexpower} 
shows the power levels of curtailment at 0\% flexibility and flexibility activation at flexibility levels of 25\% and 100\%. 
\textcolor{black}{An additional numerical result showing the activated flexibility in comparison to the load served by the substation is shown in Fig. \ref{fig:flexact}.}
\textcolor{black}{Note that ramp up flexibility (analogous to generation curtailment) activation is happening during the day when solar generation exceeds the load seen from the substation, leading to localized congestion issues due to reverse power flow and over voltage. Similarly, the ramp-down flexibility (analogous to load curtailment) activation is happening during the night and evening when the load is high, leading to congestion issues due to under-voltage and exceeding line loading limits. }\\

\subsubsection{Key takeaways}
The key takeaways are as follows:
\begin{itemize}
    \item curtailment of load and generation forms the lower bound of temporal and location flexibility needs of a DN,
    \item flexibility pricing affects resource activation. This indicates the combinatorial aspect of multiple possible solutions of RDOPF,
    \item From Table \ref{tab:resourceactiva}, we observe that RDOPF minimizes the resource activation cost and reduces cumulative line losses, 
    \item Observe from Table \ref{tab:resourceactiva}, RDOPF tends to reduce permissible under voltages, i.e. $V \in (V_{\min}, 1- \Delta_{\text{perm}}]$, more than permissible overvoltages, i.e. $V \in [1+ \Delta_{\text{perm}}, V_{\max})$.
\end{itemize}

\subsection{\textbf{Case study 3}: Effect of load reactive power compensation}
The impact of reactive power in correcting network issues has been explored widely. However, due to the already high-power factor in most low voltage distribution \textcolor{black}{systems} and high R/X ratio, the impact of reactive power flexibility is often ignored. 
This assumption of ignoring reactive compensation may not be valid with a large amount of distributed generation interfaced to the low voltage network via an inverter \cite{hashmi2020arbitrage}.

\subsubsection{Performance indices}
In this numerical study, we quantify the impact of reactive compensation for a high R/X network with a varying average power factor of the distribution network loads.
The fraction of reactive power is calculated as $Q =P \times \tan(\arccos(\text{power factor}))$. 
 The distribution network power factor takes the value in the set $\{ 0.98, 0.95, 0.9, 0.85, 0.8\}$.
 \textcolor{black}{Power factors of loads are assumed to be lagging as traditionally DN load is more inductive than capacitive.}

The metrics used to quantify the impact of reactive power flexibility are:
\begin{itemize}
    \item \textit{Reduction in active power flexibility}: Consider the total cumulative flexible energy activation by RDOPF given as $X_{noQ}$, where $X_{noQ} = |\text{load curtailed energy}| + |\text{generation curtailed energy}|$ + $|\text{ramp down curtailed energy}|$ + $|\text{ramp up curtailed energy}|$ calculated in absence of reactive compensation and at different levels of flexibility and varying power factor of the load.
    \textcolor{black}{The absolute value function is used so that the ramp-up and ramp-down energy do not compensate for each other.}
    Consider, the total absolute active energy in presence of reactive compensation is given as $Y_Q$.
    Reduction in active energy requirement is given as
    \begin{equation}
        P_{\text{reduction}} (\%) = 100 \times \frac{X_{noQ} - Y_Q}{X_{noQ}}.
    \end{equation}
    \item \textit{Profit due to reactive compensation}: Previously we defined $P_{\text{reduction}}$ which measures the reduction in active energy needs of the network. Due to the reduction in active energy, the objective function value also decreases, thus profit due to reactive compensation increases.
    The profit is given as
    \begin{equation}
        \text{Profit}_{\text{reactive}} (\%) = \frac{\texttt{Obj}_{\text{NoQ}} - \texttt{Obj}_{\text{WithQ}}}{\texttt{Obj}_{\text{NoQ}}} \times 100, 
    \end{equation}
    where $\texttt{Obj}_{\text{NoQ}} $ and $\texttt{Obj}_{\text{WithQ}} $ denotes objective function values without and with reactive compensation.
\end{itemize}

\begin{figure}[!htbp]
	\center
	\includegraphics[width=0.8\linewidth]{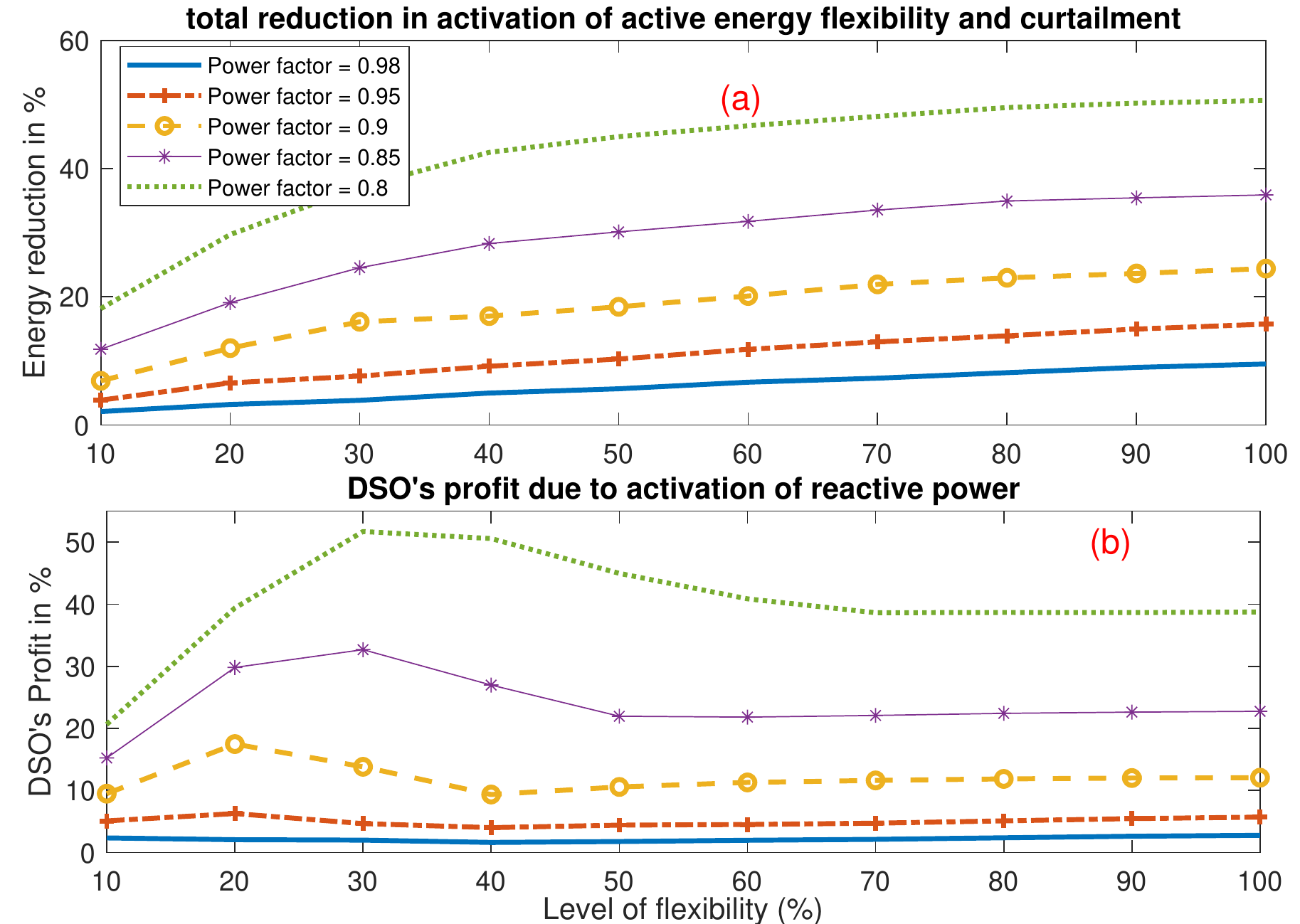}
	\vspace{-5pt}
	\caption{\small{Effect of reactive power activation on P flexibility needs and profit \textcolor{black}{of DSO. DSO's profit refers to the difference between avoided active power flexibility procurement cost, which the DSO otherwise would have to buy in the absence of Q flexibility, and the cost for Q flexibility procured.}}}
	\label{fig:c3reactive}
\end{figure}


\subsubsection{Numerical results}
The numerical results are shown in Figure \ref{fig:c3reactive}. 
Figure \ref{fig:c3reactive}(a) shows $P_{\text{reduction}}$ 
and 
Figure \ref{fig:c3reactive}(b) shows $\text{Profit}_{\text{reactive}}$.
It is clear that the active power needs of a DN reduce with reactive power compensation. For DN with a power factor of 0.8, up to 50.6\% lower active energy is needed for flexibility and for curtailment.
The profit increases much faster for lower levels of flexibility than for higher levels of flexibility. 
Note that in Figure \ref{fig:c3reactive}(b) the profit is reducing as the value of the objective function in the presence of a high degree of active power flexibility significantly decreases, refer to Table \ref{tab:resourceactiva}. Due to the reduction in objective value, the value of reactive compensation gradually deteriorates and saturates to a level.

The DSO profit for opting for reactive compensation with DN mean power factor of 0.8 exceeds 51\%.
The proportional increase in active energy reduction and profit with different levels of the power factor shows the significance of DN load power factor and reactive compensation.
For networks with low values of DN power factor, will require reactive compensation planning.
\textcolor{black}{In future work, we will quantify the importance of reactive power flexibility in low-voltage DNs. This is mainly governed by DN parameters, load variation, and DN power factor fluctuations. }

\pagebreak


\section{Conclusion}
\label{section6}

\textcolor{black}{This work explores different modelling perspectives of the novel} network-state driven resource activation mechanism. 
The flexibility activation signal considers the nodal voltage sensitivities calculated a priori and nodal voltage and projected thermal loading of a node without any flexibility or curtailable resources. 
The flexibility activation signal has similarities with the duals variables \textcolor{black}{associated with power balance constraints} of the OPF. 
The resource dispatch optimal power flow (RDOPF) is a nonlinear optimization problem that is convexified using second order cone relaxations for the voltage constraint.
\textcolor{black}{In this paper, three modelling choices of the RDOPF are quantified.}
The first numerical illustration compares the output of SOC relaxed RDOPF and AC RDOPF without any relaxations.
It is observed that penalizing DN losses reduces the gap between SOC and AC RDOPF. However, penalizing losses leads to increased activation of flexible resources.
A Pareto optimal mechanism is proposed to tune the penalty factor for the loss component of the objective function of RDOPF.
Secondly, we provide a mechanism to quantify the location and temporal flexibility needs of a DN.
\textcolor{black}{In the framework of decision support for the DSO, such temporal and locational quantification can be utilized  for the operation and planning of flexibility in a DN.}
Finally, we quantify the impact of reactive power flexibility activation on the active power flexibility needs of a DN, while also reducing the cumulative resource activation cost. We observe that as the DN power factor deteriorates, the impact of reactive power flexibility increases. For DN power factor of 0.8, reactive power flexibility can reduce operational cost and the amount of active power flexibility needs by more than 50\%.

\textcolor{black}{In future works, we will extend RDOPF for an unbalanced DN with dynamic nodal voltage sensitivities and parameter uncertainty. Further, additional exploration is required for setting up droop slopes for calculating the flexibility activation signal.}

\textcolor{black}{Further, of the accuracy of the relaxations should be assessed in a statistically relevant way, considering distribution networks with different topologies and of different sizes. As such, future work will focus on creating a methodical framework to allow such investigation.}

\pagebreak

\bibliographystyle{IEEEtran}
\bibliography{referenceOpf}

\pagebreak

    \begin{IEEEbiography}[{\includegraphics[width=.8in,height=1in,clip,keepaspectratio]{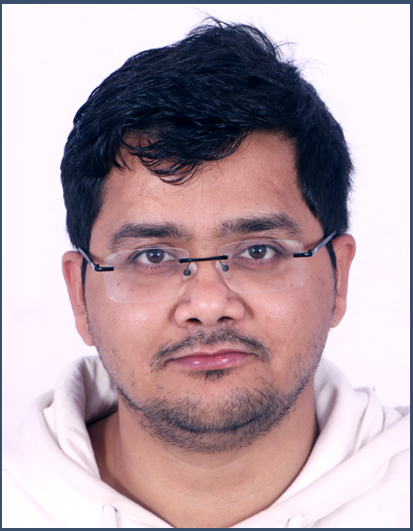}}]{Md Umar Hashmi}
    is a senior postdoctoral researcher at KU Leuven and EnergyVille in Belgium. He
       completed his PhD at \'Ecole Normale Sup\'erieure and INRIA, Paris France in December 2019. 
       He also worked for Eaton Corporation as a Controls Engineer and for Eirgrid in Dublin as Senior Engineer. 
       He completed his master's and bachelor's degree from the Indian Institute of Technology Bombay in 2012 and Aligarh Muslim University in 2010, respectively. 
       His research interests include electrical power systems, smart grids, renewable integration, data analytics, control, and optimization algorithm development.
    \end{IEEEbiography}
  \vspace{-20pt}
    \begin{IEEEbiography}[{\includegraphics[width=.8in,height=1in,clip,keepaspectratio]{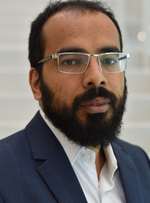}}]{Arpan Koirala}
        (Student Member, IEEE) received the master's degree under the Erasmus Mundus Master Course in Sustainable Transportation and Electrical Power Systems from the consortium of the University of Oviedo, Gijón, Spain, La Sapienza, Rome, Italy, University of Nottingham, Nottingham, U.K., and Polytechnic Institute of Coimbra, Coimbra, Portugal, in 2018. In 2023, he obtained his PhD, from the KU Leuven., Belgium. His research interests include stochastic programming, optimization, distribution system modelling, energy market, and high-scale integration of renewable resources. 
    \end{IEEEbiography}
\vspace{-34pt}
    \begin{IEEEbiography}[{\includegraphics[width=.8in,height=1in,clip,keepaspectratio]{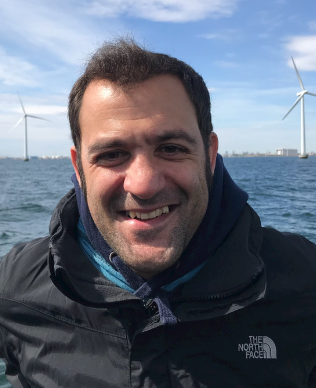}}]{Hakan Ergun}
   born in 16.02.1983 in Leoben Austria, has obtained his degree of Master of Science in Electrical Engineering at the Graz University of Technology (TU Graz) in October 2009. In February 2010, he joined the Electa Research Group at KU Leuven, Belgium, where he obtained his PhD in Electrical Engineering in January 2015. He has been a post–doctoral researcher until 2018 and is currently a research expert at KU Leuven / EnergyVille.
His main research interests are optimization methods in power system planning, power system modelling, power system reliability, electricity markets and regulation. To this date, he has worked on several national and international research projects on several aspects of power system modelling and optimization.
He has published several papers in international scientific journals and conferences. He is a senior member of IEEE and an active member of CIGRE. He has been in the local organizing committee of the IEEE EnergyCon 2016 conference in Leuven. He is a vice chair of the IEEE PES/PELS/IAS Benelux Chapter and has been the chapter chair between 2017 and 2018. He is an associate editor with the Canadian Journal of Electrical and Computer Engineering."
    \end{IEEEbiography}

    \begin{IEEEbiography}[{\includegraphics[width=.8in,height=1in,clip,keepaspectratio]{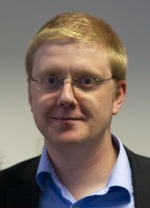}}]{Dirk Van Hertem}
     graduated as an M.Eng. in 2001 from the KHK, Geel, Belgium and as an M.Sc. in Electrical Engineering from the KU Leuven, Belgium, in 2003. In 2009, he obtained his PhD, also from KU Leuven. In 2010, Dirk Van Hertem was a member of EPS group at the Royal Institute of Technology (KTH), in Stockholm. Since spring 2011, he has been back at the University of Leuven, where he is currently professor and member of the ELECTA division. His special fields of interest are decision support for grid operators, power system operation and control in systems with FACTS and HVDC and building the transmission system of the future, including offshore grids and the supergrid concept. The research activities of Prof. Van Hertem are all part of the EnergyVille research center, where he leads the Electrical Networks activities.  Dr. Van Hertem is an active member of both IEEE (PES and IAS) and Cigré. 
    \end{IEEEbiography}

\end{document}

%% file: fig_ref.tex
\begin{figure}[!htbp]
    \centering
    
        \begin{subfigure}[b]{0.22\textwidth} 
	     \tikzsetnextfilename{fig1a}
	    \begin{tikzpicture}[line width=0.8pt, =\tikzscale]
	\definecolor{gg}{RGB}{204,255,153}
	\definecolor{orang}{RGB}{255,192,182}
	\draw[draw = none, fill = gg] (0.0,0.0) rectangle (2,-1.5);
	\draw[draw = none, fill = orang] (0.0,0.0) rectangle (-2,1.5);
	\draw[->, line width = 1pt] (-2.1, 0) -- (2.1,0);
	\draw[->, line width = 1pt] (0, -1.6) -- (0,1.75);
	\draw[line width = 0.8pt] (-2, 1.4) -- (-1.8,1.4);
	\draw[line width = 0.8pt] (-1.8, 1.4) -- (-1,0);
	\draw[line width = 0.8pt] (-1, 0) -- (1,0);
	\draw[line width = 0.8pt] (1, 0) -- (1.8,-1.4);
	\draw[line width = 0.8pt] (1.8, -1.4) -- (2,-1.4);
	\draw[line width = 0.6pt, dotted] (-1.8, 1.4) -- (0,1.4);
	\draw[line width = 0.6pt, dotted] (-1.8, 1.4) -- (-1.8,0);
	\draw[line width = 0.6pt, dotted] (1.8, -1.4) -- (0,-1.4);
	\draw[line width = 0.6pt, dotted] (1.8, -1.4) -- (1.8,0);
	\draw (1.7,0.4) node [anchor=center] {$V_{\max}$};
	\draw (-0.8,1.15) node [anchor=center] {\small{$VC^{\max}_{i,P}$}};
	\draw (0.8,-1.15) node [anchor=center] {\small{$-VC^{\max}_{i,P}$}};
	\draw (-1.8,-0.3) node [anchor=center] {$V_{\min}$};
	\draw (0.0,0.4) node [anchor=west] {\footnotesize{$\Delta V_{\text{perm}}$}};
	\draw[<->, line width = 0.5pt] (0, 0.1) -- (1,0.1);
	\draw (0,1.3) node [anchor=west] {Voltage FSA};
	\draw (0,0.9) node [anchor=west] {component};
\end{tikzpicture}
        \caption[case_1]{Voltage component}    
        \label{fig:fsacomponent1}
    \end{subfigure}
        \begin{subfigure}[b]{0.22\textwidth} 
	    \tikzsetnextfilename{fig1b}
	    \begin{tikzpicture}[line width=0.8pt, =\tikzscale]
	\definecolor{gg}{RGB}{204,255,153}
	\definecolor{orang}{RGB}{255,192,182}
	\draw[draw = none, fill = gg] (0.0,0.0) rectangle (2,1.5);
	\draw[draw = none, fill = orang] (0.0,0.0) rectangle (-2,-1.5);
	\draw[->, line width = 1pt] (-2.1, 0) -- (2.1,0);
	\draw[->, line width = 1pt] (0, -1.6) -- (0,1.75);
	\draw[line width = 0.8pt] (-2, -1.4) -- (-1.8,-1.4);
	\draw[line width = 0.8pt] (-1.8, -1.4) -- (-1,0);
	\draw[line width = 0.8pt] (-1, 0) -- (1,0);
	\draw[line width = 0.8pt] (1, 0) -- (1.8,1.4);
	\draw[line width = 0.8pt] (1.8, 1.4) -- (2,1.4);
	\draw[line width = 0.6pt, dotted] (-1.8, -1.4) -- (0,-1.4);
	\draw[line width = 0.6pt, dotted] (-1.8, -1.4) -- (-1.8,0);
	\draw[line width = 0.6pt, dotted] (1.8, 1.4) -- (0,1.4);
	\draw[line width = 0.6pt, dotted] (1.8, 1.4) -- (1.8,0);
	\draw (1.8,-0.3) node [anchor=center] {$100\%$};
	\draw (0.8,1.15) node [anchor=center] {\small{$TC^{\max}_{i,P}$}};
	\draw (-0.8,-1.15) node [anchor=center] {\small{$-TC^{\max}_{i,P}$}};
	\draw (-1.7,0.3) node [anchor=center] {$-100\%$};
	\draw (0.0,0.4) node [anchor=west] {\footnotesize{$\Delta T_{\text{perm}}$}};
	\draw[<->, line width = 0.5pt] (0, 0.1) -- (1,0.1);
	\draw (0,1.7) node [anchor=east] {Thermal FSA};
	\draw (0,1.3) node [anchor=east] {component};
\end{tikzpicture}
        \caption[case_2]{Thermal component }    
        \label{fig:fsacomponent2}
    \end{subfigure}

   \vspace{-13pt}
    \caption{\small{Convention, components for active (P) and reactive (Q) power FAS design based on DN states. 
    $\Delta T_{\text{perm}}$ and $\Delta V_{\text{perm}}$ are the thresholds set for thermal and voltage corrections.
    The ramp down P needs are denoted by orange color, and vice versa. Ramp down P flexibility reduces net load. 
    In case of Q, the region shaded in orange denotes the need for capacitive flexibility and green denotes inductive Q. }}
    \label{fig:fasdesign}
\end{figure}